\documentclass[twocolumn]{aastex63}
\usepackage{epsfig}
\usepackage{amsmath, amsthm, amssymb}
\usepackage{microtype}
\usepackage{float}
\usepackage{verbatim}
\usepackage{wrapfig}
\usepackage{graphicx}
\usepackage[caption = false]{subfig}
\usepackage{multirow}
\usepackage{hhline}
\usepackage{booktabs}
\usepackage{array}
\usepackage{bm}
\usepackage[normalem]{ulem}
\usepackage{changepage}

\usepackage{rotating}

\newcommand{\numBWsTotal}{41}
\newcommand{\numBWsconfirmed}{37}
\newcommand{\numTids}{7}
\newcommand{\numXrays}{20}

\shorttitle{Black Widow Millisecond Pulsars}
\shortauthors{Swihart et al.}

\begin{document}

\title{A New Flaring Black Widow Candidate and Demographics of Black Widow Millisecond Pulsars in the Galactic Field}

\correspondingauthor{Samuel J. Swihart}
\email{samuel.swihart.ctr@nrl.navy.mil}

\author{Samuel J. Swihart}
\affiliation{National Research Council Research Associate, National Academy of Sciences, Washington, DC 20001, USA,\\ resident at U.S. Naval Research Laboratory, Washington, DC 20375, USA}
\author{Jay Strader}
\affiliation{Center for Data Intensive and Time Domain Astronomy, Department of Physics and Astronomy,\\ Michigan State University, East Lansing, MI 48824, USA}
\author{Laura Chomiuk}
\affiliation{Center for Data Intensive and Time Domain Astronomy, Department of Physics and Astronomy,\\ Michigan State University, East Lansing, MI 48824, USA}
\author{Elias Aydi}
\affiliation{Center for Data Intensive and Time Domain Astronomy, Department of Physics and Astronomy,\\ Michigan State University, East Lansing, MI 48824, USA}
\author{Kirill V. Sokolovsky}
\affiliation{Center for Data Intensive and Time Domain Astronomy, Department of Physics and Astronomy,\\ Michigan State University, East Lansing, MI 48824, USA}
\affiliation{Sternberg Astronomical Institute, Moscow State University, Universitetskii pr. 13, 119992 Moscow, Russia}
\author{Paul S. Ray}
\affiliation{Space Science Division, U.S. Naval Research Laboratory, Washington, DC 20375, USA}
\author{Matthew Kerr}
\affiliation{Space Science Division, U.S. Naval Research Laboratory, Washington, DC 20375, USA}

\begin{abstract}
We present the discovery of a new optical/X-ray source likely associated with the \emph{Fermi} $\gamma$-ray source 4FGL J1408.6--2917. Its high-amplitude periodic optical variability, large spectroscopic radial velocity semi-amplitude, evidence for optical emission lines and flaring, and X-ray properties together imply the source is probably a new black widow millisecond pulsar binary. We compile the properties of the 41 confirmed and suspected field black widows, finding a median secondary mass of $0.027\pm0.003\,M_{\odot}$. Considered jointly with the more massive redback millisecond pulsar binaries, we find that the ``spider" companion mass distribution remains strongly bimodal, with essentially zero systems having companion masses between $\sim0.07-0.1\,M_{\odot}$. X-ray emission from black widows is typically softer and less luminous than in redbacks, consistent with less efficient particle acceleration in the intrabinary shock in black widows, excepting a few systems that appear to have more efficient ``redback-like" shocks. Together black widows and redbacks dominate the census of the fastest-spinning field millisecond pulsars in binaries with known companion types, making up $\gtrsim$80\% of systems with $P_{\rm{spin}}<2\,\rm{ms}$. Similar to redbacks, the neutron star masses in black widows appear on average significantly larger than the canonical $1.4\,M_{\odot}$, and many of the highest-mass neutron stars claimed to date are black widows with $M_{\rm{NS}}\gtrsim2.1\,M_{\odot}$. Both of these observations are consistent with 
an evolutionary picture where spider millisecond pulsars emerge from short orbital period progenitors that had a lengthy period of mass transfer initiated while the companion was on the main sequence, leading to fast spins and high masses.

\vspace{10mm}
\end{abstract}

\section{Introduction}
In the standard ``recycling'' scenario for forming millisecond pulsars (MSPs), a rotating neutron star is spun up to rapid spin periods through accretion from a binary companion \citep{Alpar82}. Once mass transfer ends, the resulting binary system consists of a MSP with a low-mass white dwarf companion in a relatively wide ($P\gtrsim2$ d) orbit \citep{Tauris06}. Prior to the launch of the \emph{Fermi}-LAT satellite in 2008, nearly all of the known MSP binaries in the Galactic field (i.e., outside of globular clusters) were in these end-stage systems.

Since 2008, dozens of field MSPs with short-period ($P\lesssim1$ d), low mass, non-degenerate companions have been discovered, many by conducting multiwavelength follow-up of \emph{Fermi} $\gamma$-ray sources. These compact binaries are typically referred to as ``spiders'' due to the evaporative effects the high-energy pulsar wind has on the companion, and are further subdivided into either ``black widows'' or ``redbacks'' depending on whether the companion is less or more massive than $\sim0.1\,M_{\odot}$ \citep{Roberts13}.

In these spider binaries, the radio pulsar is often obscured or eclipsed due to ionized material being blown from the companion by the relativistic pulsar wind, making it difficult to find pulsations using typical radio search techniques, especially at binary phases when the companion lies between Earth and the MSP \citep[e.g.,][Swihart et al.~in prep.]{Camilo16,Cromartie16,Deneva16,Corongiu21}. However, nearly all MSPs appear to be $\gamma$-ray emitters \citep{Abdo13}, and since these high-energy photons flow through the diffuse companion material unimpeded, multiwavelength follow-up of unidentified \emph{Fermi} sources continues to be a successful way to discover and characterize these binaries \citep[e.g.,][]{Nieder20,Li21,Swihart21,Ray22,Swihart22}.

Optical and near-IR light curves of these spider systems show characteristic variability modulated on the orbital period of the binary. For most redbacks, this modulation has two peaks per orbital cycle (i.e., ellipsoidal variations) corresponding to when we are the viewing the maximum surface area of the tidally distorted companion as it orbits the MSP \citep[e.g.,][]{Salvetti15,Bellm16,Li16}. In the black widows (and some redbacks), irradiation from the energetic pulsar wind dominates over tidal distortion effects, causing one face of the tidally locked companion to be heated to a much higher temperature than the side facing away from the MSP \citep[e.g.,][]{Breton13,Romani16,Linares18,Draghis19,Swihart19}. In these cases only one broad peak in the light curve is observed per orbit when viewing this heated face at companion superior conjunction, along with one minimum when we see the unheated ``nightside'' of the companion as it crosses between Earth and the MSP.

The tidal synchronization timescale for typical spider MSPs is short ($\lesssim$ few Myr), so the orbital period equals the rotation period of the secondary. This rapid rotation along with irradiative heating of the photosphere from the pulsar wind can drive strong winds from the secondary \citep[e.g.,][]{Morin12,Romani15}. At the intersection between the stellar wind and the pulsar wind, an intrabinary shock can form that emits nonthermal X-rays, often modulated on the orbital period of the binary \citep[e.g.,][]{Bogdanov14,Romani16b,Wadiasingh17,Noori18}.

Balmer and Helium optical emission lines can also be attributed to the stellar winds directly, or in combination with the intrabinary shock that serves as a natural location for a region with high enough temperature to produce H$\alpha$ and He I photons \citep{Romani15,Halpern17b,Swihart18,Strader19}. The combination of all these effects can cause significant flaring and variability on rapid timescales, both in the overall brightness of the system and in the strength and location of emission lines \citep[e.g.,][]{Cho18,Halpern22}.

In addition to the significant radio eclipses seen in the spider binaries, these systems also show other differences from the traditional population of MSPs, including higher neutron star masses and faster spins \citep{Strader19}. Three of the most massive neutron stars known to date that have relatively well-constrained neutron star mass measurements are in spider binaries, each with neutron star mass estimates that exceed $2\,M_{\odot}$ \citep{Kandel20,Romani21,Romani22b}. We note however that the component mass estimates in many spider MSPs often rely on binary inclination measurements derived from fitting irradiation-dominated light curves, which are difficult to model as described in the following sections.

In this paper, we present the discovery of a new compact binary associated with the \emph{Fermi} source 4FGL J1408.6--2917 as part of our ongoing program following up X-ray/optical matches within unassociated \emph{Fermi} regions. The X-ray properties and variable optical light and radial velocity curves strongly suggest this is a new black widow MSP binary. We introduce this source as part of a compilation of the confirmed and candidate field black widows. This census of black widow MSPs, totalling~\numBWsTotal~systems as of mid-2022, is compared to the closely related redbacks, showing strong bimodality in some of their observed and intrinsic multiwavelength properties, despite (possibly) sharing a similar evolutionary path.

\begin{figure*}[ht!]
\begin{center}
	\includegraphics[width=1.0\linewidth]{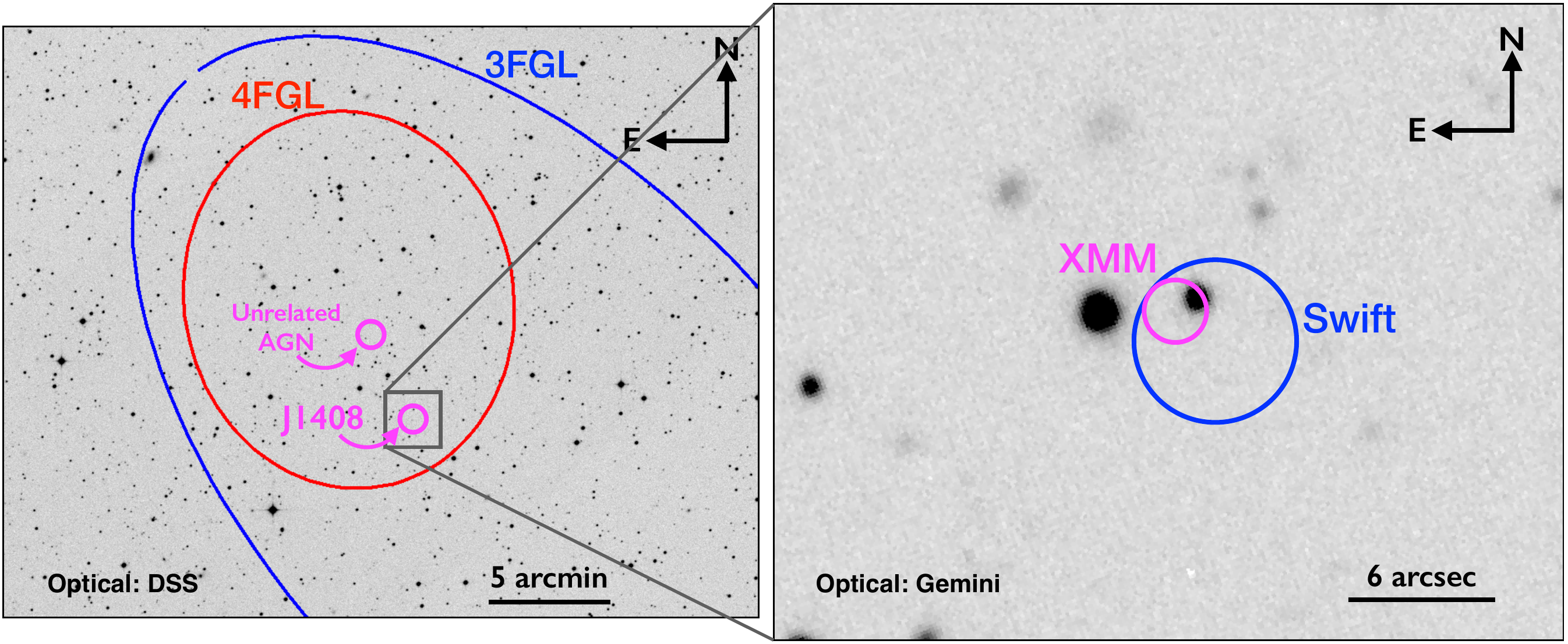}
    \caption{Left: Optical Digital Sky Survey image of the field showing the 95\% error ellipses from the 3FGL (blue) and 4FGL (red) catalogs corresponding to the $\gamma$-ray source 4FGL J1408.6--2917. The relative position of the two X-ray sources in the 4FGL region (see Sec.~\ref{sec:X-rays}) are marked with magenta circles. The likely black widow counterpart to the \emph{Fermi} source is labeled as J1408. Right: Gemini-South/GMOS $i'$ image zoomed in on the position of J1408. The \emph{Swift} and \emph{XMM} X-ray positions and their 90\% confidence regions are shown with blue and magenta circles, respectively. The variable optical source discussed in Sec.~\ref{sec:optLC} is spatially coincident with the X-rays and is the likely companion in a black widow MSP binary.}
\end{center}
\label{fig:finder_fig}
\end{figure*}

\section{Observations/Data}
\label{sec:obs}

\subsection{$\gamma$-rays}
\label{sec:gammarays}
The gamma-ray source associated with a new MSP binary candidate presented here for the first time is listed in the third incremental data release of the fourth full catalog of \emph{Fermi}-LAT sources \citep[4FGL-DR3;][]{Fermi22} as 4FGL J1408.6--2917. Based on twelve years of survey data, the source has an overall detection significance of 11.3$\sigma$ in the 0.1--100 GeV energy range, and has been detected in the previous 1FGL, 2FGL, and 3FGL catalogs. The 4FGL 95\% confidence error region lies entirely within the 3FGL region and is $\sim$80\% smaller in area (Figure~\ref{fig:finder_fig}). The source shows no significant variability over the lifetime of \emph{Fermi}, consistent with most other MSPs. The $\gamma$-ray spectrum shows marginal evidence for curvature; while some confirmed MSPs show strongly curved GeV spectra, others show a similar lack of strong spectral curvature \citep[e.g.,][]{Strader15}.

\subsection{X-rays}
\label{sec:X-rays}

\subsubsection{Swift}
\label{sec:Swiftobs}
4FGL J1408.6--2917 was targeted with \emph{Swift}/XRT on seven epochs between 2019 Jun to Sep. In an automated analysis of these data \citep{Stroh13}, one X-ray source is present within the 4FGL ellipse and this source also matches to a faint ($G=20.2$) \emph{Gaia} source, which we later determined was a high-amplitude optical variable (Sec.~\ref{sec:optcounterparts}), motivating additional follow-up. For the remainder of the paper, we refer to this optical/X-ray source as J1408.

We also obtained a $\sim$1.4 ksec \emph{Swift} ToO observation of the region on 2021 Feb 25. Including the data described above, the total on-source XRT exposure time used in the following analysis is 8.0 ksec.

We used the online \emph{Swift}-XRT analysis tools\footnote{https://www.swift.ac.uk/user\_objects/} \citep{Evans20} to analyze these data, finding J1408 with a ICRS position (R.A., Dec.) of (14:08:26.77, --29:22:22.7) and a 90\% positional uncertainty of 3.9\arcsec~(Figure~\ref{fig:finder_fig}). With a net count rate of only $\sim3\times10^{\rm{-3}}$ ct s$^{\rm{-1}}$, there were not enough source counts to assess variability with these data.

This analysis also revealed a second \emph{Swift} source, closer to the center of the 4FGL region, but fainter than J1408, with only $\sim$ 8 net counts, at a (R.A., Dec.) position of (14:08:35.40, --29:18:57.4) and a 90\% positional uncertainty of 6.4\arcsec. We classify this source as a distant active galactic nucleus (AGN) unrelated to the 4FGL $\gamma$-ray source, and defer additional discussion of this source to Appendix \ref{sec:AGN}.

\vspace{7.5mm}

\subsubsection{XMM}
\label{sec:XMMdesrip}
We obtained a ToO observation centered on the 4FGL J1408.6--2917 region with \emph{XMM-Newton}/EPIC on 2021 Jul 23. We reprocessed these data using standard tasks in the Science Analysis System (SAS, v20.0.0).

We found no evidence of strong, extended background flares, resulting in a total on-source exposure time of $\sim$24 ksec. Standard flagging and selection criteria were used for the pn and MOS cameras, respectively, as described in the online analysis threads\footnote{https://www.cosmos.esa.int/web/xmm-newton/sas-threads}.

For the spectral analysis, we extracted individual spectra from each EPIC camera using a circular source extraction radius of 25\arcsec~and local background regions three times larger. Individual pn, MOS1, and MOS2 spectra were then combined using \texttt{epicspeccombine}\footnote{https://www.cosmos.esa.int/web/xmm-newton/sas-thread-epic-merging}. We grouped the final combined spectrum into a minimum of 30 counts per bin and fit the resulting background-subtracted spectrum using \texttt{XSPEC} v12.12.1 \citep{Arnaud96}. The best position of the \emph{XMM} source associated with J1408 is overlaid onto an optical image in the right panel of Figure~\ref{fig:finder_fig}. 

As part of compiling the multiwavelength properties of the field black widows, we found three confirmed systems with archival \emph{XMM} data from observations taken in 2016 (PI: M. Roberts). These data were reduced in the same manner as described above and represent the first X-ray analyses of these systems. The results are presented in Section~\ref{sec:census}.\\

\subsection{Optical Counterpart}
\label{sec:optcounterparts}
There is one optical source matching the \emph{Swift} and \emph{XMM} X-ray positions of J1408, and it is listed in \emph{Gaia} DR3 \citep{GaiaDR3} with brightness $G=20.21 \pm 0.06$ mag and a ICRS position of (R.A., Dec.) = (14:08:26.789, --29:22:21.21). No parallax or proper motion information is available in the current \emph{Gaia} data release. This source is also listed in Pan-STARRS DR2 with a brightness of $i'=20.87 \pm 0.04$. No other optical sources are present within 4.4\arcsec~of J1408 down to a limiting magnitude of $i'\sim22.5$. This is confirmed with our deep Gemini exposures (Sec.~\ref{sec:Gemdata}).

The \emph{Gaia} photometric uncertainty is large for an isolated star at this brightness, implying a variable source \citep{Andrew21,Mowlavi21}. The association between this variable optical source and the X-ray source are confirmed by our spectroscopy and photometry presented below.

\subsubsection{SOAR Imaging and Spectroscopy}
\label{sec:SOARdata}
We obtained spectroscopy of J1408 with the red camera of the Goodman Spectrograph \citep{Clemens04} on the SOAR telescope over several epochs from 2021 Feb 18 to 2021 Aug 6, with 11 usable spectra obtained over 7 different nights. In all cases we used a 400 l mm$^{-1}$ grating with wavelength coverage from $\sim 3900$ to $7850~\AA$. Depending on the seeing, we used either a 0.95\arcsec~or 1.2\arcsec~longslit, giving full-width at half-maximum (FWHM) spectral resolutions of $\sim 5.8$\,\AA~or 7.3\,\AA, respectively. Exposure times per spectrum ranged from 1200 s to 1800 s. Due to the low resolution of the spectra, smearing due to the motion of the star during the exposure has only a marginal effect on the final data. All spectra were reduced and optimally extracted in the standard manner using IRAF \citep{Tody86}.

We also performed imaging observations with SOAR/Goodman during three nights on 2021 Feb 19, Mar 1, and Mar 3 centered on the position of J1408. On each night we took a series of exposures using the SDSS $i'$ filter. On Feb 19, exposures were 300 sec in length, while on the March nights the exposures were 400 sec. Typical seeing was 1.1", 1.2", and 1.4" on Feb 19, Mar 1, and Mar 3, respectively.

Raw images were reduced with combinations of bias and flat fields in the usual manner using IRAF. For each nightly dataset, we performed differential aperture photometry and calibrated the instrumental magnitudes of the target using the Pan-STAARS DR2 $i'$ magnitudes of 22 nearby, non-variable comparison stars \citep[see also,][]{Swihart20}.

The most obvious feature of these light curves is the rapid brightening of the source by $>$2 mag over the course of $\sim$1 hour. The source was too faint to be detected in a subset of these images, presumably when we are viewing the ``nightside'' of the companion. We removed images where the source was undetected at $\gtrsim$1.8 sigma above the background, giving uninterrupted intervals of 0.87 hr, 1.62 hr, and 1.83 hr on Feb 19, Mar 1, and Mar 3, respectively.

The final sample of SOAR photometry consists of 11, 15, and 17 measurements on these nights. The brightness peaks at $i'\sim20.35$, and is undetectable in these images around $i'\sim22.5$.

\subsubsection{Gemini/GMOS Imaging and Spectroscopy}
\label{sec:Gemdata}
Since J1408 was too faint to measure at high signal-to-noise throughout its full orbit with SOAR, we obtained a 4.5-hr series of photometric observations using Gemini-South/GMOS as part of the Gemini Fast Turnaround Program (ID: GS-2021A-FT-112). These observations were performed on 2021 Jun 06 using the $i'$ filter with a frame time of 220 sec. The seeing varied throughout the night but was typically good, around $\sim0.7\arcsec$. We performed standard GMOS-imaging data reduction routines using \texttt{DRAGONS v2.1.1} \citep{Dragons19}. The light curve was extracted using IRAF in a similar manner to the SOAR data, and calibrated using the same 22 nearby, non-variable comparison stars.

J1408 was significantly detected in all images. The source shows a clear periodic modulation of $\sim$3.42 hrs, rising to a maximum brightness of $i'\sim20.5$ and dimming to a minimum of $i'\sim22.9$. The final Gemini photometric sample consists of 66 measurements.

We show the Gemini and SOAR data folded on the best-fit period ($P\sim3.42$ hr; see Sec.~\ref{sec:BW}) in Figure~\ref{fig:paperfig_LC}, where we have set the phase of maximum brightness to $\phi=0.75$, consistent with the radial velocity peaking at $\phi=0.5$ (i.e., pulsar phase convention). Overall, the amplitude and period of this variability are fully consistent with a black widow MSP that is being heated substantially on one side facing the pulsar.

We also obtained 7 additional optical spectra of J1408 with Gemini-South/GMOS (ID: GS-2022A-FT-202) on 2022 Apr 11 over a time range of about 3.15 hr, which is nearly a full orbital cycle (3.42 hr). These data used a 1\arcsec~slit with the R400 grating and the GG455 order-blocking filter centered at 6800 \AA, with per spectrum exposure times of 25 min. These spectra have a mean FWHM resolution of about 7.2 \AA~and cover a nominal wavelength range of $\sim 4500$--9150 \AA, though the signal-to-noise toward the blue is very poor, and the data in the wavelength range $\sim 7160$--7660 \AA~is unusable. As for the SOAR/Goodman data, we reduced and optimally extracted these spectra using IRAF.

\subsection{Radio Pulsar Search Data}
We obtained a short series of pulsar search observations with the Robert C. Byrd Green Bank Telescope on 2021 Jun 14 in an effort to detect the suspected radio pulsar (Project ID: GBT21A-428). The first pointing lasted $\sim$30 min using the Prime Focus 820 MHz receiver, immediately followed by a $\sim$35 min pointing in S-band centered at 2.2 GHz.

\section{Results for J1408: A Flaring Black Widow}
\label{sec:BW}

\subsection{X-ray Flux and Spectrum}
\label{sec:xray_results}

\subsubsection{Swift}
To fit the \emph{Swift} X-ray spectrum of J1408, we used \texttt{XSPEC} (v12.12.1) to fit a simple absorbed power-law model and assumed \citet{Wilms00} abundances. The neutral hydrogen column density was held fixed to the Galactic value of $3.4\times10^{20}$ cm$^{-2}$
\citep{HI4PI}. The best-fit photon index is $\Gamma=2.4^{+1.3}_{-0.9}$ (here and throughout the paper, uncertainties on the X-ray properties are quoted at the 90\% confidence level), and the unabsorbed 0.5--10 keV X-ray flux is $F_{X} = (8^{+56}_{-5})\times10^{-14}$ erg s$^{\rm{-1}}$ cm$^{\rm{-2}}$ .

\subsubsection{XMM}
The deeper \emph{XMM} dataset allows for a much more precise determination of the X-ray spectrum, with a
EPIC count rate of $0.010\pm0.002$ ct s$^{\rm{-1}}$ and a total of $\sim$244 net counts. We fit a similar power-law model as with the \emph{Swift} data (\texttt{TBabs*powerlaw}). Leaving $N_{H}$ free in the fit resulted in very low values, so we held $N_{H}$ fixed to the Galactic value. The best-fitting photon index from these data is $\Gamma=1.7^{+0.6}_{-0.5}$ with an unabsorbed 0.5--10 keV flux of $(2.4^{+1.6}_{-1.1})\times10^{-14}$ erg s$^{\rm{-1}}$ cm$^{\rm{-2}}$. Although this median flux value is $\sim$70\% less than the flux from the \emph{Swift} data, the two values are consistent within their uncertainties. The \emph{XMM} data provides $\gtrsim10\times$~more counts than the \emph{Swift} dataset so we assume the results from the \emph{XMM} analysis for the remainder of the paper.

In an attempt to constrain the orbital variability, we divided the 0.2--10 keV background-subtracted light curve into four time bins, each spanning about 6 ksec. Within the large uncertainties, there is no evidence for significant variability, though we note that even factor of $\sim 2$ variability could not be detected at high significance given the low count rate.

Some black widow MSPs show dramatic orbital variability associated with the intrabinary shock \citep[e.g.,][]{Huang12}, with the largest effects occurring in systems that are more edge-on \citep[e.g.,][]{Romani16b, Wadiasingh17}. A more face-on orbit would weaken the orbital effects of the intrabinary shock emission, so the fact that we see little evidence for X-ray variability may be indicative of a relatively low orbital inclination angle. A somewhat face-on orbit is supported by both the light curve models and our spectroscopy, so deeper X-ray observations to detect variability at the low levels expected here are well-motivated.

\subsection{Pulsar Search}
For the Green Bank Telescope pulsar search data obtained on 2021 Jun 14, we excised RFI and searched the data for periodic signals using typical procedures in \texttt{PRESTO} v3.0.1 \citep{Ransom11}. The short orbital period of the system relative to our observation duration may cause ``smearing'' of the pulsar signal in the Fourier domain. We therefore implemented the ``jerk'' search feature in \texttt{PRESTO} (i.e., \texttt{-wmax}) to improve the search sensitivity \citep{Anderson18}. We searched both the full datasets as well as $\sim$15 min subsets of each pointing with \texttt{accelsearch} up to \texttt{-zmax 200} and with and without \texttt{-wmax 600}.

These searches did not produce any clear detections of a radio pulsar. Using the orbital period and ephemerides presented below, the 820 MHz observation occurred during the orbital phase range $0.69 < \phi < 0.81$, and $0.88 < \phi < 1.03$ for the S-band pointing. These orbital phases (especially during the 820 MHz observation) should be favorable for minimizing absorption of the radio pulsar by ionized material associated with the likely black widow companion, though we note that these represent a single search epoch.

Pulsar non-detections have also been found for some other likely spider MSPs: systems which have strong optical and/or X-ray evidence for the presence of a pulsar, but in which no radio pulsar has been detected despite numerous efforts \citep[e.g.,][]{Swihart20,Corbet22,Halpern22}. One possibility is that even at higher frequencies, in some spider MSPs the material lost from the companion may eclipse the pulsar to an even greater degree than the 5--15\% of a typical spider (e.g., \citealt{1991ApJ...380..557R,1996ApJ...465L.119S,2020MNRAS.494.2948P}). Another possibility is that in a subset of systems, the radio pulsar beams do not sweep over our line of sight. Given the strong evidence that J1408 is indeed a black widow, additional pulsar search observations would be valuable.

\subsection{Optical Light Curve}
\label{sec:optLC}
\begin{figure*}[t!]
\begin{center}
	\includegraphics[width=0.7\linewidth]{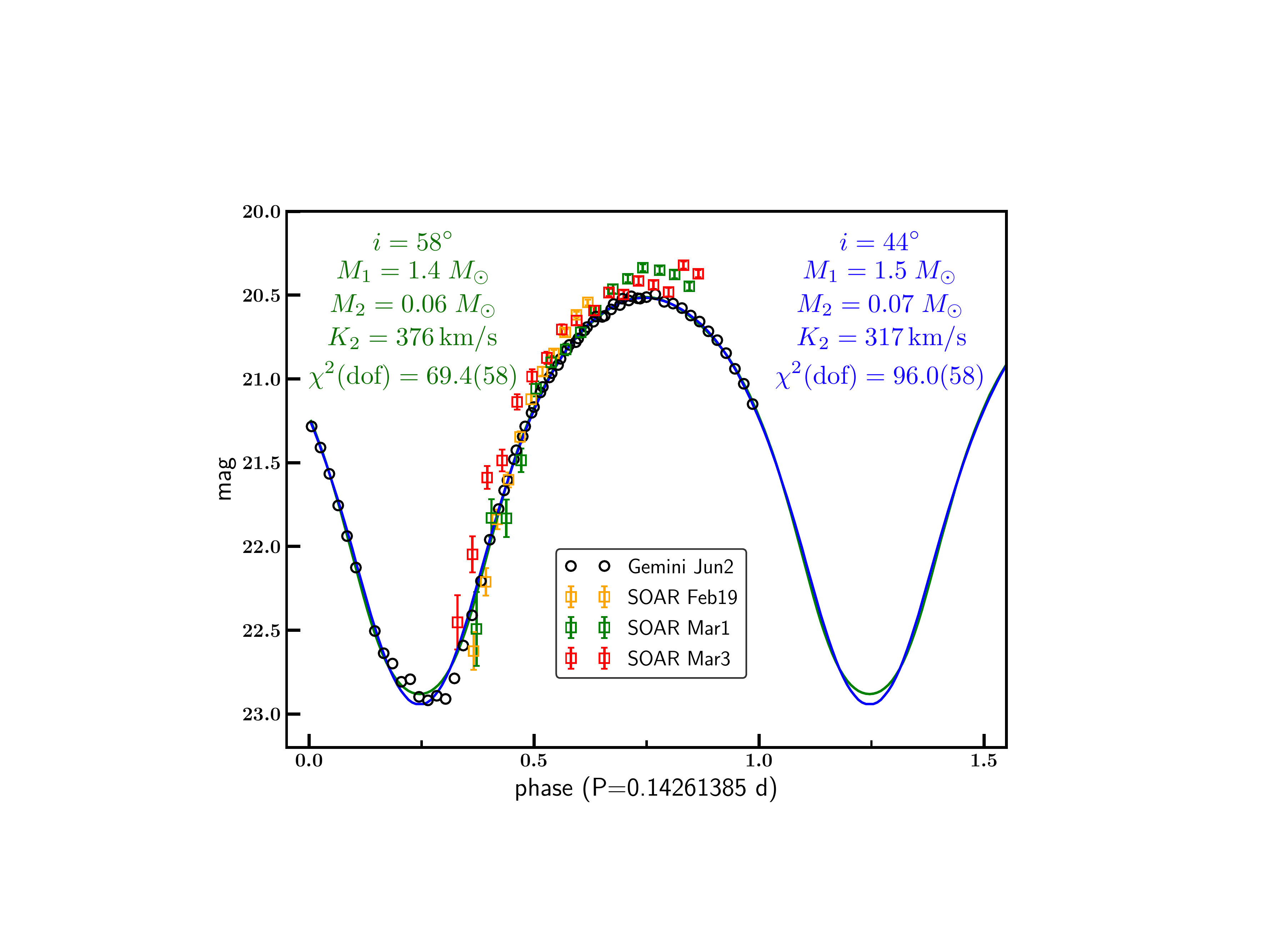}
    \caption{SOAR and Gemini $i'$-band photometry of the likely black widow companion to 4FGL J1408.6--2917. The data are folded on the best fit period, $P\sim3.42$ hr. The ephemeris is set such that $\phi=0.75$ at maximum brightness, the presumed superior conjunction of the heated companion. While the Gemini data are relatively smooth and well-behaved, the SOAR data shows evidence of flares that vary from night to night, especially after $\phi\sim0.65$ (Sec.~\ref{sec:optLC}). Two representative light curve models are shown (green and blue lines/text) that only differ significantly in the orbital phase range $\phi\sim0.2-0.35$ (see Sec.~\ref{sec:LCmodels}).
    }
\label{fig:paperfig_LC}
\end{center}
\end{figure*}

The uninterrupted Gemini photometry shows a strong periodic signal at $P\sim3.42$ hr. When combined with the SOAR data, the period that best agrees with the full photometric dataset is $P=0.14261385 \pm 0.0000015$ d, which we take as the best period. When folded on this period (Figure~\ref{fig:paperfig_LC}), the photometry shows a single, broad, large-amplitude peak, consistent with the light curves of other black widow binaries. In this context, the broad peak corresponds to when we are viewing the hot irradiated face of the companion that is being heated substantially by the pulsar on its tidally locked ``dayside,'' whereas the narrower minimum corresponds to when we are viewing the much cooler ``nightside'' of the companion when it lies between Earth and the suspected neutron star primary on its orbit (i.e., companion inferior conjunction).

Overall, the Gemini light curve is rather smooth, showing no evidence for flaring. This is in contrast to the SOAR data, which shows what appears to be minor flares as the source rises to peak brightness. This is especially apparent in the 2021 Mar 3 data where the final two points are significantly brighter ($\gtrsim$ 0.1 mag) than the preceding measurements at a time when the light curve is expected to have turned over towards fainter values. The 2021 Mar 3 photometry also seems to be slightly brighter overall than the other datasets throughout the orbit. This is also apparent during the end of the 2021 Mar 1 epoch, where the photometry agrees well with the Gemini data until $\phi\sim0.65$ where the source rises quickly to be $\sim$0.1--0.15 mag brighter than the Gemini data at corresponding phases. A similar effect is observed at the end of the 2021 Feb 19 SOAR epoch starting just after $\phi\sim0.5$.

The observing conditions were similar during each SOAR night, and we used the exact same set of comparison stars in our analysis, so we conclude the variations we see in the photometry are real. These flaring optical light curves are not uncommon in spider binaries, with similar phenomenology seen in the black widow candidate 4FGL J0935.3+0901 \citep{Halpern22} and in a few confirmed and candidate redbacks \citep{Cho18,Halpern22b}.

The exact mechanism causing these flares in spider systems is still unclear, but is likely caused by a combination of variable intrabinary shock emission and anisotropic heating due to ducting of the pulsar wind or shock emission along the companion magnetic field lines \citep[e.g.,][]{Romani15,Sanchez17}. In this context, rapid structural variations in the heating mechanism or a variable companion magnetic field can provide natural explanations for fast flaring activity on timescales of days to weeks \citep{Cho18}. If the heating occurs high enough in the companion's atmosphere that it creates a thin layer of ionised material in the chromosphere, these effects may also contribute to the rapid appearance and disappearance of H and He emission lines in the spectra. 

These flaring effects can operate simultaneously with other processes that give rise to slower variations in the continuum emission from the companion, which can cause large heating asymmetries in the light curves, such as hot/cold spots on the stellar surface, or diffusion and convection within the photosphere due to the large temperature asymmetry on the companion surface \citep[e.g.,][]{Swihart19,Kandel20,Voisin20}.

We also observe flaring in the optical spectra, which we describe in detail in Sec.~\ref{sec:spec_results}.

\subsubsection{Light Curve Modeling}
\label{sec:LCmodels}
We used the Eclipsing Light Curve code \citep[\texttt{ELC};][]{Orosz00} to provide an initial model of the optical light curve of the system. While this model includes the effect of direct heating of the companion, it does not include other potentially relevant physical effects such as zonal heat flow that have been shown to improve light-curve fits for spider companions (e.g., \citealt{Kandel20}). At present we only have a rather limited, single-band light curve for J1408, so more sophisticated modeling is not yet justified.

The SOAR photometry has incomplete phase coverage, large uncertainties when $i'\gtrsim21.5$, and suffers from irregular flaring. We therefore only model the Gemini light curve, which covers the full orbit, is more precise, and shows no evidence for significant flaring. 

Absent a pulsar timing solution or constraints on the binary mass ratio, we assumed a primary mass consistent with a neutron star ($\sim1.4-2.0\,M_{\odot}$) and fit for the binary inclination $i$, Roche lobe filling factor of the companion $f_{2}$, base (nightside) temperature of the companion $T_{2}$, the isotropic irradiating luminosity from the pulsar (we characterize this quantity as the maximum dayside temperature of the heated secondary $T_{\rm{day}}$), and the binary mass ratio $q=M_{2}/M_{\rm{NS}}$. We also assumed the orbital period from the photometry. 

In general, a wide range of models fit the data equally well, all with component masses fully consistent with a black widow MSP binary. The statistical uncertainties on the Gemini photometry are very small, so after finding a range of good-fitting models, we inflated the photometric errors by a factor of $2.6\times$ so that the total reduced $\chi^2$ of the final model was closer to 1.0, a method commonly used to model the light curves of spider MSPs \citep[e.g.,][]{Linares18,Swihart20}. We note that the best-fitting model values are not sensitive to the exact multiplicative value used to inflate the photometric errors (we did not fine-tune this value so our final models have reduced $\chi^2$ slightly greater than 1.0). Within the uncertainties, the overall results are identical for models with and without inflating the photometric errors.

Overall, the best-fitting model has $i=58.0^{\circ}$, $f_{2}=0.95$, $T_{2} = 3837$ K, $T_{\rm{day}} = 8797$ K, with a primary mass of $M_{1}=1.40\,M_{\odot}$ and secondary mass of $M_{2}=0.058\,M_{\odot}$. This is a good fit statistically with a reduced $\chi^{2}$ ($\chi^{2}_{\rm{red}}$ = $\chi^{2}$/dof) of 69.4/58 = 1.2. However the $K_{2}$ value associated with this fit is 376 km s$^{\rm{-1}}$, larger than the value derived from our SOAR and Gemini spectroscopy (Sec.~\ref{sec:spec_results}).

There are a range of other models that have similarly good statistical fits but that agree better with our spectroscopy-derived $K_{2}$ value. For example, a model with $i=44.0^{\circ}$, $f_{2}=0.98$, $T_{2} = 2895$ K, $T_{\rm{day}} = 6246$ K, and $M_{1}=1.5\,M_{\odot}$, $M_{2}=0.07\,M_{\odot}$ results in a fit with $\chi^{2}_{\rm{red}}$ = 1.6, but with $K_{2}=317$ km s$^{\rm{-1}}$.

Setting $T_{2} = 2400$ K, the approximate mean nightside temperature of the nine black widows fit by \citet{Draghis19}, returns a best-fitting model with $\chi^{2}_{\rm{red}}$ = 1.2. 
The model parameters for this fit are $i=55.0^{\circ}$, $f_{2}=0.98$, $T_{\rm{day}} = 3750$ K, $M_{1}=1.4\,M_{\odot}$, $M_{2}=0.020\,M_{\odot}$, and $K_{2}=369$ km s$^{\rm{-1}}$.

All of these models produce nearly identical fits to the data between binary phases $\phi=0.4-1.2$. The largest differences between models occur near minimum brightness ($\phi\sim0.2-0.35$) where the fit residuals are highest, especially near $\phi\sim0.3$. Figure~\ref{fig:paperfig_LC} shows the small differences in the model light curves near these phases.

Given the wide range of generally well-fitting models, the single photometric filter of the dataset, and the incomplete inclusion of physical effects in the models, we do not quote formal uncertainties on the fitted and derived model values. Broadly, we find that the light curve is consistent with that of a near-Roche-lobe-filling companion to a neutron star with an intermediate inclination $i \sim 44-58^{\circ}$ and potentially a companion mass that is on the high side for black widows ($\sim 0.05-0.07\,M_{\odot}$). We defer a closer comparison with the spectroscopic results to Section \ref{sec:spec_results}.

\vspace{9mm}

\begin{deluxetable}{lcc}[!t]
\label{table:spec_obs}
\tablecaption{Summary of SOAR \& Gemini Spectroscopic Observations}
\tablehead{
\colhead{Date} & \colhead{Binary Phase} & \colhead{Emission Lines?}
}
\startdata
\bf{SOAR} & &\\
\hline
2021 Feb 18 & 0.90 & \checkmark\\
2021 Feb 20 & 0.75 & --\\
2021 Jun 09 & 0.07 & --\\
2021 Jul 16 & 0.92 & --\\
2021 Jul 16 & 0.07 & --\\
2021 Jul 28 & 0.71 & \checkmark\\
2021 Jul 28 & 0.84 & \checkmark\\
2021 Jul 29 & 0.99 & \checkmark\\
2021 Jul 30 & 0.80 & \checkmark\\
2021 Jul 31 & 0.98 & \checkmark\\
2021 Aug 06 & 0.76 & --\\
\hline
\bf{Gemini} & &\\
 \hline
2022 Apr 11 & 0.76 & --\\
2022 Apr 11 & 0.89 & --\\
2022 Apr 11 & 0.02 & \checkmark\\
2022 Apr 11 & 0.15 & \checkmark\\
2022 Apr 11 & 0.30 & \checkmark\\
2022 Apr 11 & 0.43 & \checkmark\\
2022 Apr 11 & 0.56 & \checkmark
\enddata
\end{deluxetable}

\subsection{Optical Spectroscopy}
\label{sec:spec_results}

\subsubsection{SOAR Spectroscopy}
Due to the extreme faintness of J1408 at optical minimum ($i\sim23$), most of the SOAR spectra were obtained at phases closer to optical maximum ($\phi =0.75$), from $\phi = 0.71$ to 0.99, with two additional spectra at $\phi \sim 0.07$. This means that the orbital phase coverage of the SOAR spectra is generally poor.

\begin{figure*}[t!]
\begin{center}
	\includegraphics[width=\linewidth]{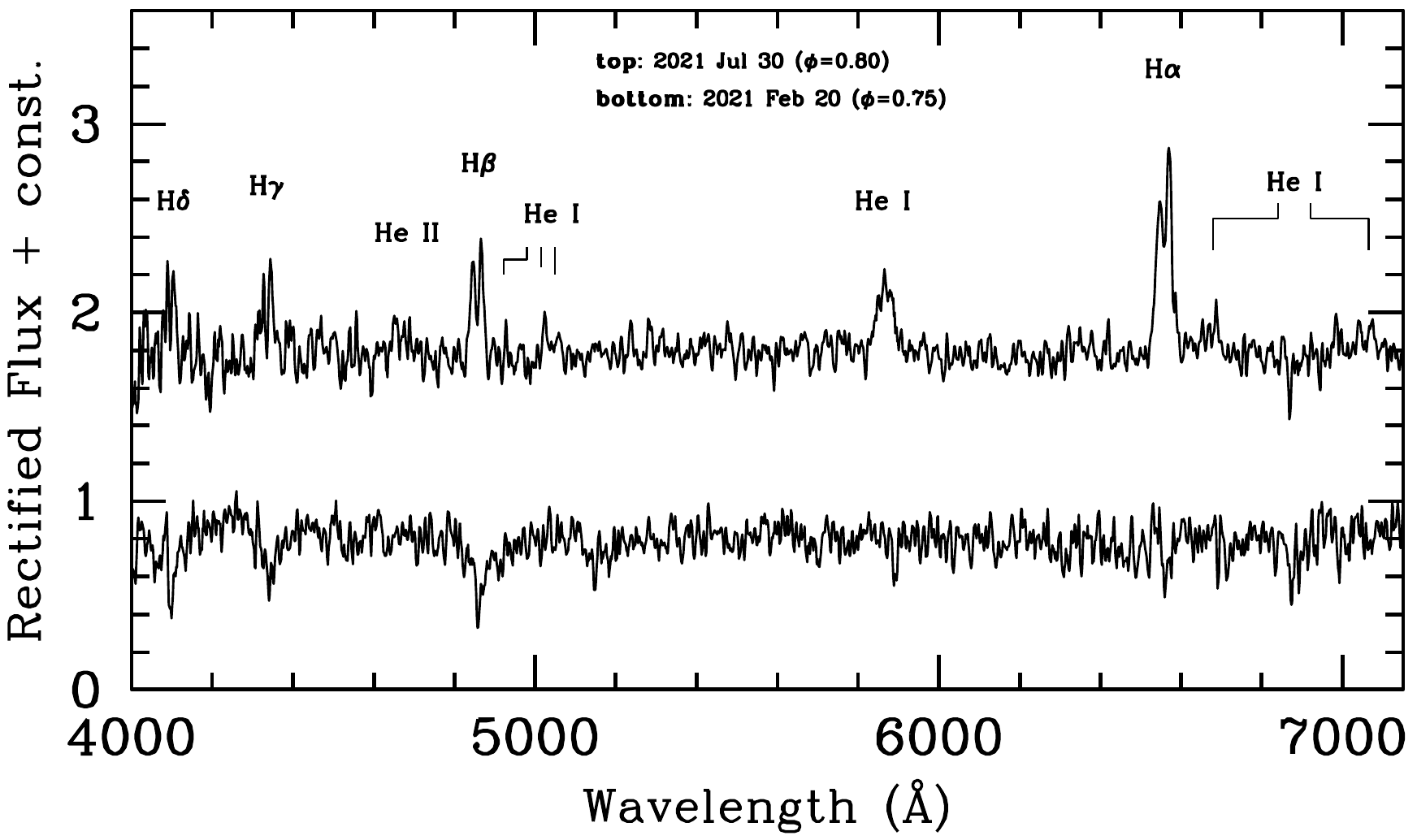}
    \caption{Two SOAR spectra of J1408 taken at similar orbital phase on different days, showing the variations observed at different epochs. In the bottom spectrum, taken on 2021 Feb 20 ($\phi = 0.75$), no emission lines are seen, but instead primarily hydrogen Balmer lines in absorption. In the top spectrum, from 2021 Jul 30 ($\phi = 0.80$), the dominant features are instead broad/double-peaked H and He emission lines.}
    \label{fig:comp_j1408}
\end{center}
\end{figure*}

Of the 11 SOAR spectra, 6 clearly show broad luminous emission lines of H and in some cases He, while in the other spectra no emission is seen. Given the narrow range of phases in the SOAR data ($\phi = 0.71-1.07$), it is immediately clear that the presence of the emission lines is not primarily related to the orbital phase. The Gemini spectra (Sec.~\ref{sec:gem_spec}) provide additional support for this conclusion (Table~\ref{table:spec_obs}).

In the first SOAR spectrum of J1408, obtained on 2021 Feb 18, both  H (H$\alpha$ and H$\beta$) and \ion{He}{1} (5875 and 7065 \AA) emission is observed. The FWHM of H$\alpha$ is quite broad at $\sim 1770$ km s$^{-1}$. No emission lines were visible in four subsequent spectra taken from 2021 Feb 20 to Jul 16, nor in the final SOAR spectrum obtained on 2021 Aug 6.

Emission lines are clearly visible in all three SOAR spectra obtained on 2021 Jul 28/29, which were the next data taken after 2021 Jul 16. In the first two spectra ($\phi=0.71$ and 0.84) the emission lines are clearly double-peaked, with H$\alpha$ peak separations of $\sim 1425$ and 1150 km s$^{-1}$, respectively. Fitting a Gaussian to the full double-peaked line shapes gave FWHMs of $\sim 2820$ and 2290 km s$^{-1}$, respectively. In the subsequent spectrum ($\phi=0.99$) the emission lines are broad but no longer double-peaked, with an H$\alpha$ FWHM of $\sim 1510$ km s$^{-1}$. A similar evolution occurs in the pair of SOAR spectra obtained on the night of 2021 Jul 30/31, where emission lines including H$\alpha$ are double-peaked in the first spectrum ($\phi=0.80$) but not in the second ($\phi=0.98$). 

The signal-to-noise in the continuum of the SOAR spectra ranges from poor to modest. Metal-line absorption at Mg$b$ is visible in a few spectra, even including several on 2021 Jul 28/29 and 30/31 where emission lines are also present. No spectra showing broad M star absorption lines are seen, but this may be primarily due to the poor orbital phase coverage. In one spectrum (2021 Feb 20) no metal lines are seen, but H$\beta$, H$\gamma$, and H$\delta$ are instead observed in absorption (H$\alpha$ is marginal, and likely partially filled in by emission), suggesting a warmer surface temperature at this epoch.

Figure~\ref{fig:comp_j1408} compares the emission line spectrum of 2021 Jul 30 to the absorption line spectrum of 2021 Feb 20, which were each taken during similar binary phases.

\subsubsection{Gemini Spectroscopy}
\label{sec:gem_spec}

The seven Gemini/GMOS spectra were taken in series, covering nearly a full orbit of J1408. Continuum emission was present for the first two spectra in the series ($\phi = 0.76$, 0.89). Absorption lines associated with the donor were seen clearly in the first spectrum and marginally in the second, and there were no emission lines visible in either spectrum. The continuum was very faint in the third ($\phi = 0.02$) and fifth ($\phi = 0.30$) spectra and was undetectable in the fourth ($\phi = 0.15$). This general trend in the continuum flux is expected, as the mean photometric maximum is expected around $\phi = 0.75$ and the minimum at $\phi=0.25$. 

Over the timespan of the $\phi = 0.02$ to 0.30 spectra, strong, broad emission lines grew in prominence. In the $\phi = 0.02$ spectrum only H$\alpha$ is visible, but it is joined by marginally detectable \ion{He}{1} 5785 \AA\ in the $\phi = 0.15$ and $\phi = 0.30$ spectra. In the final two spectra ($\phi = 0.43$, 0.56), the continuum re-brightened toward the expected maximum at $\phi =0.75$. However, the emission lines did not fade, but continued to increase in flux. In the $\phi=0.56$ spectrum, in addition to H$\alpha$ and \ion{He}{1} 5785 \AA, H$\beta$ and part of the Paschen series in the red, as well as \ion{He}{1} 6678 and 7065 \AA, were all observed clearly. No photospheric absorption lines were seen in these final two spectra despite the re-emergence of the continuum. 

The H$\alpha$ FWHM increased from $\sim 640$ km s$^{-1}$ at $\phi = 0.02$ to $\sim 1660$ km s$^{-1}$ for $\phi = 0.30$. The FWHM in the final spectrum ($\phi =0.56$) is similar to this highest value, but is substantially lower in the penultimate ($\phi=0.43$) spectrum at $\sim 1160$ km s$^{-1}$, suggesting a non-monotonic trend. The emission lines were well resolved, but not double-peaked, in all the Gemini/GMOS spectra in which they were apparent.

\subsubsection{Emission Lines and Their Origin}

Between the SOAR and Gemini spectroscopic datasets, broad emission lines were observed at all orbital phases in at least one epoch. However, at most phases, J1408 shows emission in some epochs, but not in others, suggesting there is no simple relationship between orbital phase and the presence of emission (the exception is $\phi=0.2$--0.6, which only are covered by a single Gemini epoch).
The FWHM of the emission lines also show no clear relationship to orbital phase. 

Instead, it is likely, that as observed in some other black widows and redbacks with sufficiently extensive optical spectroscopy, that the emission lines are primarily associated with an intrabinary shock between the stellar wind of the secondary and the pulsar wind of the primary. This shock is not constant in time, but instead shows flaring activity.

\begin{figure}[t!]
\begin{center}
	\includegraphics[width=\linewidth]{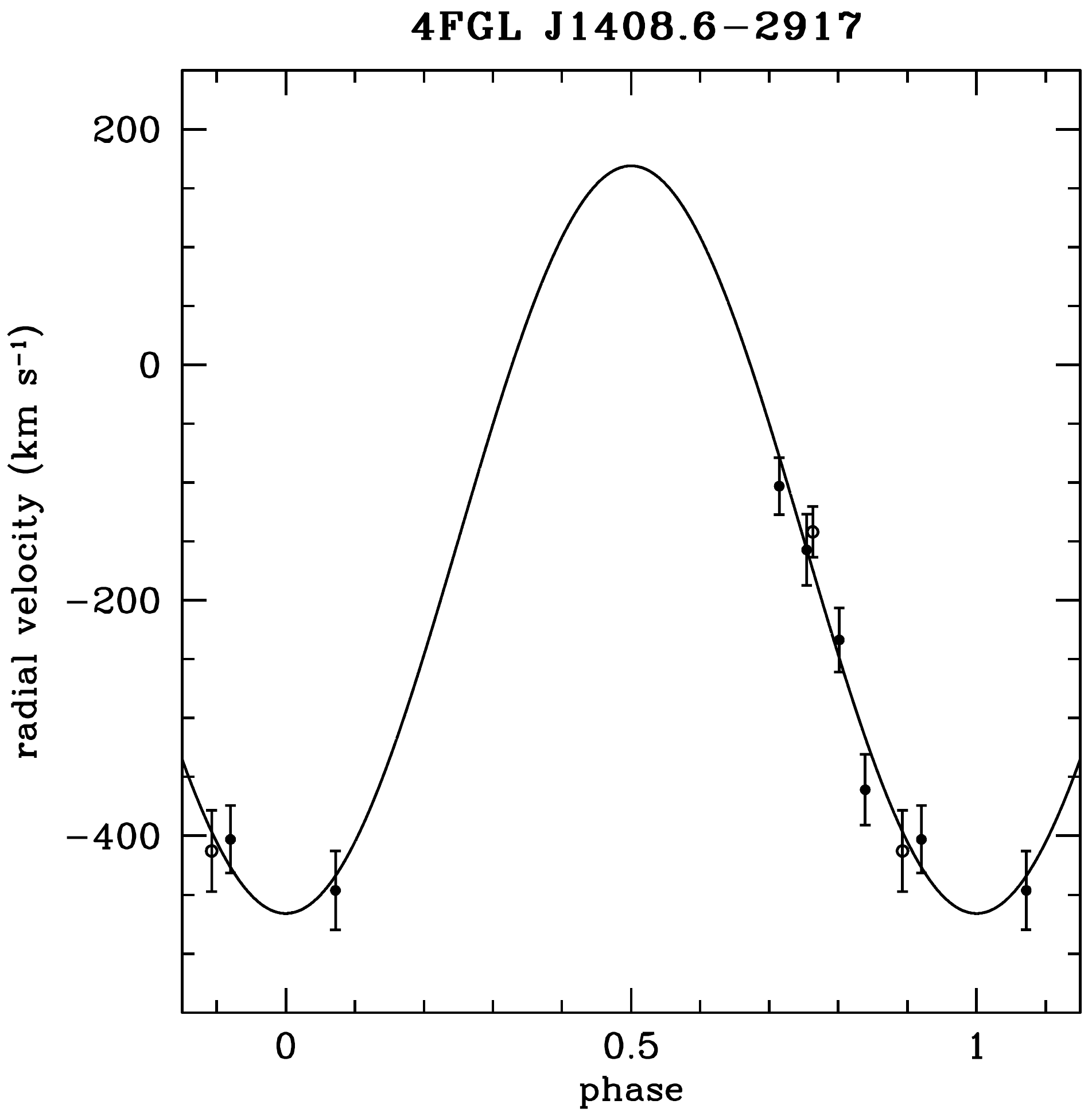}
    \caption{Circular Keplerian fit to the SOAR (filled circles) and Gemini (open circles) radial velocities.}
    \label{fig:rv_j1408}
\end{center}
\end{figure}

\subsubsection{Absorption Lines, Radial Velocities, and Orbital Solution}

For the SOAR spectra that showed evidence of a continuum and metal absorption lines, we attempted to derive radial velocities of the companion star through cross-correlation around the Mg$b$ region with bright template stars of early to mid-K. For the 2021 Feb 20 spectrum, the strongest features were Balmer lines in absorption, and a warmer template was used for cross-correlation in the regions of H$\beta$, H$\gamma$, and H$\delta$. Six of the eleven SOAR spectra yielded absorption-line radial velocities.

Only two of the seven Gemini spectra showed evidence for absorption lines. Since the continuum signal in these spectra was higher in the red than in the blue, radial velocities for these two spectra were derived through cross-correlation in the Ca triplet region.

We next fit a circular Keplerian model to these eight radial velocities. Owing to the small number of velocities and their substantial uncertainties, we fixed the orbital period and epoch of the ascending node of the pulsar to the best-fit photometric values. The latter assumption may not be correct due to asymmetric heating, but in practice does not seem obviously wrong. 

The best-fit model has $K_{2,obs} = 317\pm31$ km s$^{-1}$ and $v_{sys} = -148\pm16$ km s$^{-1}$, with the uncertainties inferred via bootstrap. Owing to the poor phase coverage of the velocities, these quantities are correlated, with a lower $K_{2,obs}$ implying a more negative $v_{sys}$. This best fit is shown in Figure \ref{fig:rv_j1408}, and has a $\chi^2$/d.o.f = 6.8/6.

Taken at face value, the mass function implied by the measured $K_{2,obs}$ and orbital period is only $0.47\pm0.14\,M_{\odot}$, suggesting a likely inclination of J1408 of $i \lesssim 45^{\circ}$. This more face-on inclination could also help explain why metal absorption lines are observed even at superior conjunction of the secondary, when in a typical black widow system this is when the warm irradiated face should dominate the spectrum.

However, there is a mild tension between this $K_{2,obs}$
and the predicted $K_{2}$ from the best-fitting light curve model (376 km s$^{-1}$, so different at the $1.7\sigma$ level). The uncertainty in $K_{2,obs}$ is larger than typical due to the poor phase coverage of the spectra for which radial velocities could be derived. Another possibility is that the correction from a center-of-light to center-of-mass $K_2$ might partially address this tension (though the corrections for black widows typically go the other direction, e.g.,  \citealt{Kandel20}).

We conclude that the kinematics of the secondary are generally consistent with expectations for a black widow companion, but that the joint spectroscopic and light curve evidence is equivocal between a typical intermediate inclination and a somewhat more face-on inclination. Future multi-band light curves and ``lucky" spectroscopy when the flaring happens to be less prominent, and/or a pulsar timing solution, can help address these uncertainties.

\section{Black Widow Census}
\label{sec:census}
\subsection{The Black Widow Sample}
We have compiled many of the multiwavelength properties of the confirmed and candidate black widows in the Galactic field 
in Tables~\ref{table:BWcatalog}--\ref{table:BWcatalog4}. Confirmed systems are defined here as those where a radio millisecond pulsar ($P_{\rm{spin}}<8\,\rm{msec}$) has been detected, and which have a companion mass $\lesssim0.1\,M_{\odot}$ (sans PSR J1908+2105, see below). As shown in Sec.~\ref{sec:compmasses}, a small change to this mass cutoff has no effect on the sample as there are essentially no spider MSPs with companion masses between $0.07-0.1\,M_{\odot}$.

There are~\numBWsconfirmed~systems that meet these criteria. Of these,~\numTids~have companions that are extremely low mass, with minimum companion masses $\lesssim0.01\,M_{\odot}$ (implying very large mass ratios, $M_{\rm{NS}}/M_{c}\gtrsim150$, assuming typical neutron star masses). It may be the case that the evolutionary paths leading to these extreme-mass-ratio systems differ from the bulk of the black widow distribution \citep[e.g.,][]{Romani16}, but given their broad similarities to the black widows we include them here.

At the bottom of the tables, we also include four candidate systems for which radio pulsations have not yet been detected, but that have strong evidence for a black widow classification based on the multiwavelength data. These are 4FGL J1408.6--2917 (this work), 4FGL J0335.0+7502 \citep{Li21}, 4FGL J0935.3+0901 \citep{Wang20, Halpern22}, and ZTF J1406+1222 \citep{Burdge22}, a recently discovered candidate black widow in a hierarchical triple. 

We do not include PSR J1908+2105 in this list of black widows. Its likely companion mass ($>0.055\,M_{\odot}$) is in a sparsely populated mass range, consistent with either the most massive black widow companions or the low-mass end of redbacks, and the extensive radio eclipses observed in this system are more characteristic of redbacks than black widows \citep{Cromartie16, Strader19}.

\subsection{Measured Properties}
\label{sec:measprops}
A vast majority of the black widows have a precise pulsar timing solution that enables tight constraints on the typical pulsar parameters, namely the spin period and its derivative ($P_{\rm{spin}}$, $\dot{P_{\rm{obs}}}$), projected semimajor axis ($a\,\rm{sin}\,i$), dispersion measure (DM), and pulsar spin-down power ($\dot{E}$). For the candidates with no pulsar detection the \emph{Gaia} position is used, while for the others the best position is taken from the ATNF database\footnote{https://www.atnf.csiro.au/research/pulsar/psrcat/} (Table~\ref{table:BWcatalog}).

For compiling distances to each source, we take a hierarchical approach. The most accurate and precise distances come from radio timing parallax measurements. For the four systems with a reliable timing parallax we adopt those distances and associated uncertainties. One source, PSR J1653--0158, has a moderately precise \emph{Gaia} parallax measurement ($\widetilde{\omega}/\sigma_{\widetilde{\omega}} \sim 2.3$), and we adopt the associated geometric distance for this source \citep{BailerJones21}.

For systems with modeled optical light curves, an estimate of the distance to the binary is possible by comparing the observed fluxes to the flux predicted by the light curve model assuming some radius and temperature for the companion \citep[e.g.,][]{Breton13,Swihart17,Draghis19}. For the spider MSPs with a significant \emph{Gaia} parallax measurement, and therefore a precise geometric distance estimate, these optical light curve derived distances have been shown to be more accurate than the DM-based distance estimates \citep{Jennings18}. For sources without parallax measurements, we adopt these optically-derived distances unless the authors in the cited reference suspect the distance is unreliable (for example, \citet{Draghis19} suspect the distance they estimate to PSR J0251+2606 may be dubious due to the incomplete light curve coverage). Lastly, for sources with no parallax or light-curve derived distance estimates, we adopt the \citet{Yao17} DM model distances since they have been shown to be more accurate than the \citet{Cordes02} model for pulsars, especially those outside the Galactic plane \citep{Jennings18}. In these cases we assume 30\% uncertainty on the DM-based distance model value. We list these distance estimates in Table~\ref{table:BWcatalog2}.

For the~\numXrays~systems that have been observed in X-rays, we list the unabsorbed 0.5--10 keV X-ray fluxes ($F_{X}$) and best-fit power-law photon index ($\Gamma$) in Table~\ref{table:BWcatalog3}. We corrected all the X-ray fluxes to this standard energy range using WebPIMMS\footnote{https://heasarc.gsfc.nasa.gov/cgi-bin/Tools/w3pimms/w3pimms.pl} assuming the best-fit flux and photon index in the reference cited \citep[see also,][]{Lee18}. For most sources, the listed $\Gamma$ assumes a simple power-law model, unless a combined thermal plus power-law model was a significantly better fit. If a combined thermal plus power-law model was comparable statistically to the simple power-law, we assumed the simpler model. For some of the brighter sources, a phase-resolved X-ray analysis was performed to determine whether the X-ray properties differ based on binary phase \citep[e.g.,][]{Kandel21}. In these cases we assumed the properties of the phase-averaged (i.e., full orbit) spectrum for consistency.

For three systems, PSR J1946--5403, PSR J2052+1219, and PSR J2115+5448, we found unpublished archival \emph{XMM} observations and analyzed these with the same procedures as described in~Sec.~\ref{sec:XMMdesrip}. To our knowledge, this work presents the first analysis of these systems in X-rays.

Using the final adopted distance and associated uncertainty, we list the 0.5--10 keV X-ray luminosity ($L_{X}$) and 0.1--100 GeV $\gamma$-ray luminosity ($L_{\gamma}$) if the source is detected with \emph{Fermi} (Table~\ref{table:BWcatalog3}).

Binary orbital periods from pulsar timing are adopted when they are available, otherwise we assume the best period from the optical photometry and/or spectroscopy (Table~\ref{table:BWcatalog4}).

\subsubsection{MSP Spin Distributions}

\begin{figure}[]
\begin{center}
	\includegraphics[width=\linewidth]{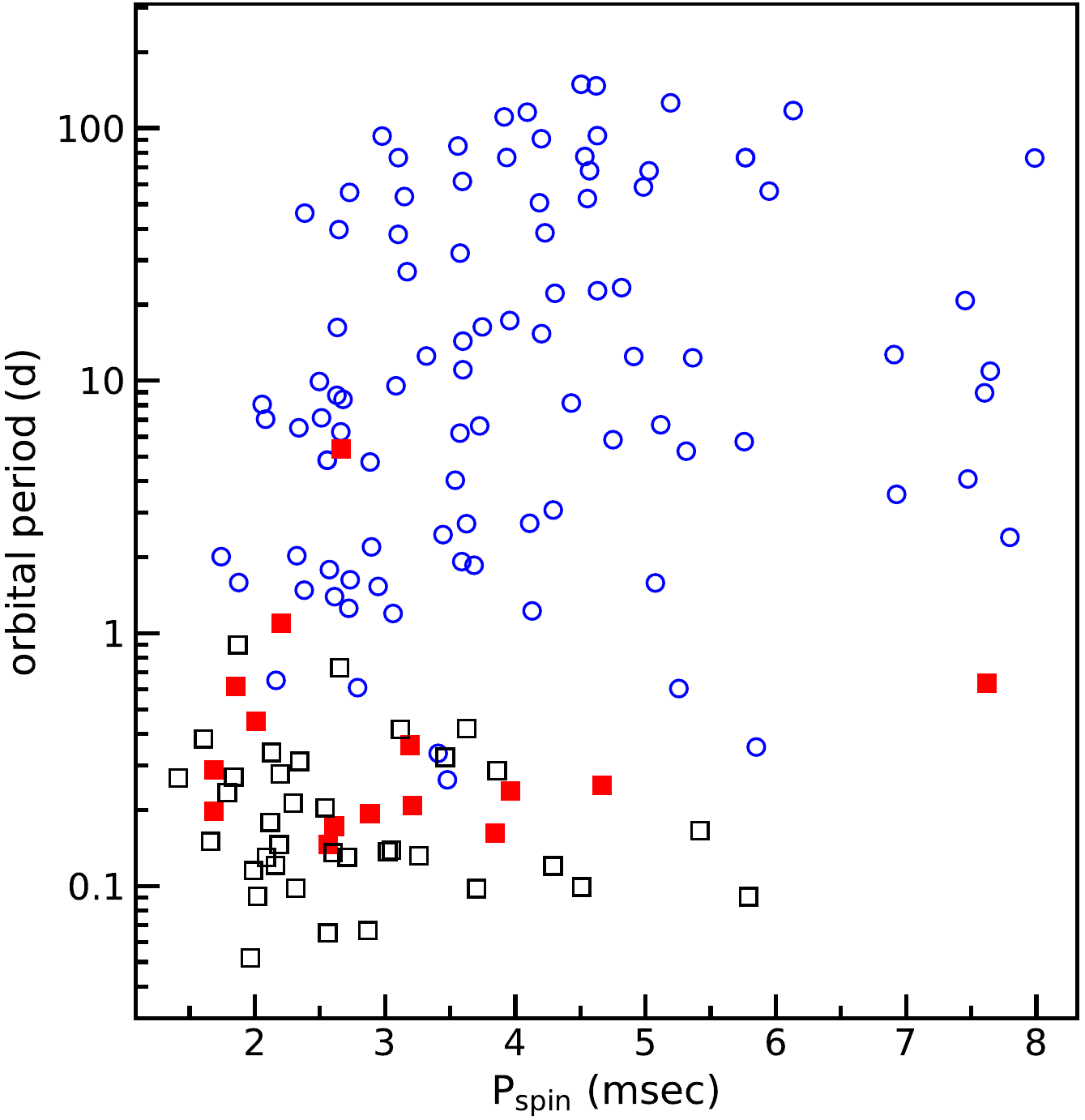}
    \caption{Spin period versus orbital period for the redbacks (red) and black widows (black) along with the field MSPs with He white dwarf companions (blue) highlighting the large fraction of spiders among the fastest spinning systems.
    }
    \label{fig:spins_fig}
\end{center}

\end{figure}

In Figure~\ref{fig:spins_fig} we show the spin period versus the orbital period for the field black widows and redbacks along with field MSP--He white dwarf binaries. This figure highlights what a large fraction of the fastest spinning MSPs are in spider binaries. Among the 34 systems with $P_{\rm{spin}} \leq 2.5\,\rm{ms}$, 24 (71\%) are confirmed spiders. At even shorter spin periods ($P_{\rm{spin}} \leq 2.0\,\rm{ms}$), 11 of 13 (85\%) are black widows/redbacks.

The short orbital period pre-MSP binaries that begin transferring mass on the main sequence will naturally mass transfer for a longer period of time relative to systems that don't start mass transfer until the companion has evolved off the main sequence \citep[e.g.,][]{Tauris99}. The spin distribution we observe in Figure~\ref{fig:spins_fig} is therefore direct confirmation of the behavior expected if the progenitors of spiders indeed had very short initial orbital periods compared to the progenitors of typical MSP--He white dwarf binaries.

\subsubsection{Masses}

Reliable constraints on the neutron star masses in black widow binaries are difficult to obtain because these estimates often rely on the binary inclination derived from modelling the optical light curves, which can be plagued by systematic uncertainties when the light from the companion is dominated by irradiation. Similar heating effects make it hard to estimate the binary mass ratio, since this often relies on an accurate measurement of the semi-amplitude of the companion radial velocity curve, which must be carefully corrected for the difference between the system's center-of-mass and its center-of-light. Given the large (and often poorly characterized) systematic uncertainties associated with these effects, we refrain from including neutron star mass estimates in this bulk catalog. However, we do explicitly list the systems for which photometry/spectroscopy exists, which allow for more accurate estimates of the component masses when coupled with a precise pulsar timing solution (Table~\ref{table:BWcatalog4}). While we recognize the uncertainties involved, we proceed with discussion of the companion mass estimates in the following section. Although we do not list the neutron star mass estimates explicitly in the catalog, we summarize some of the recent literature about neutron star masses in black widows in Sec.~\ref{sec:NSmasses}.

\begin{figure}[t!]
\begin{center}
	\includegraphics[width=\linewidth]{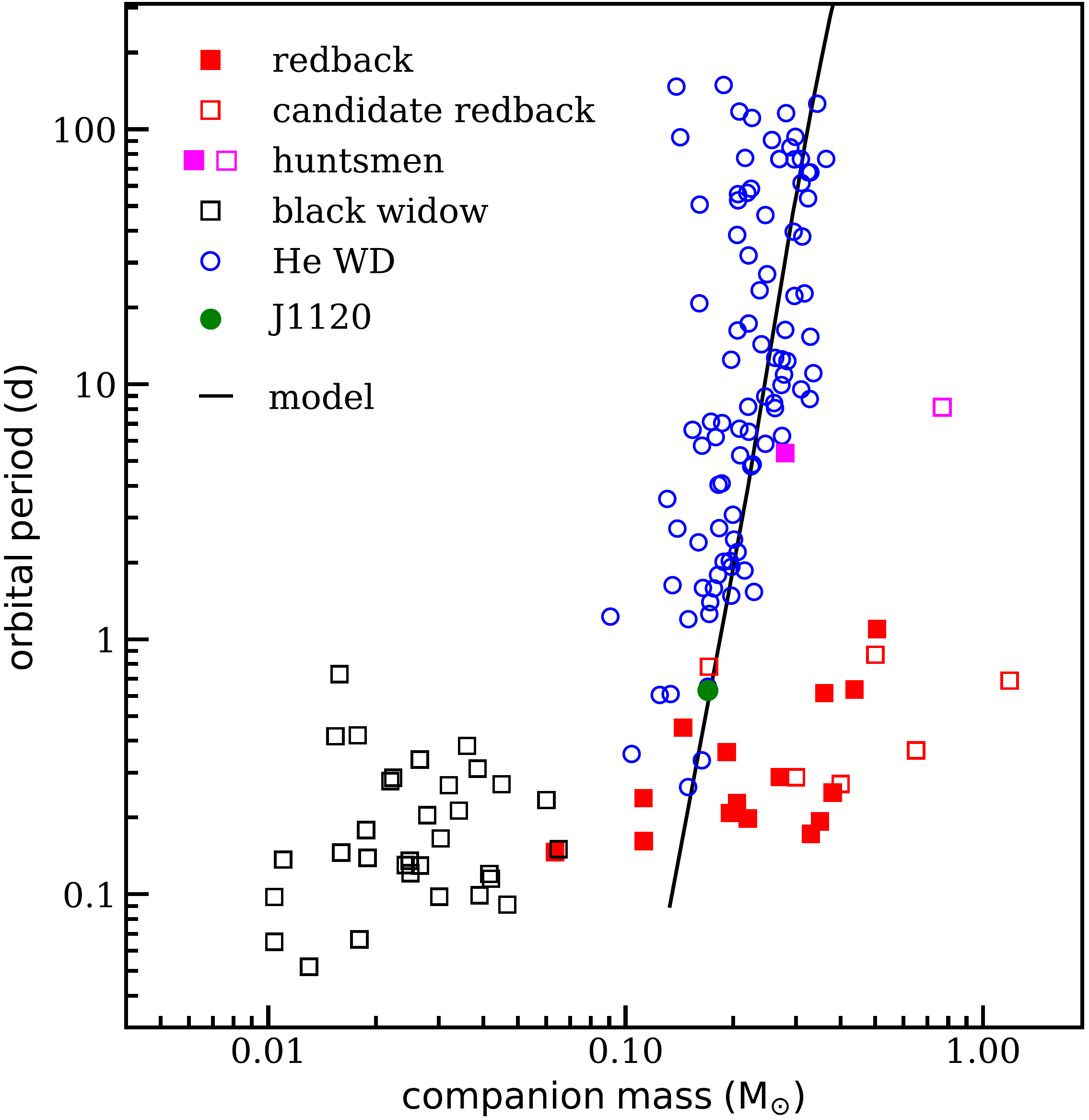}
    \caption{Median (or best-fit, if available) companion mass vs. orbital period for the field redbacks (red) and black widows (black). MSPs with He white dwarf companions are also shown (blue) along with a binary evolution model which assumes an initial secondary mass of $1.0\,M_{\odot}$ and denotes the endpoints of an ensemble of systems with varying initial period \citep{Tauris99}. The recently discovered MSP--proto white dwarf binary associated with 4FGL J1120.0--2204 \citep{Swihart22} (green circle) and the two long-period ``huntsmen'' systems with (sub)giant companions (orange) are progenitors of ``normal'' MSP--He white dwarf binaries. Despite the growing number of discoveries made in the last several years, there is still a notable absence of sources with companion masses in the range $0.07-0.1\,M_{\odot}$.
    }
    \label{fig:spiderfig}
\end{center}

\end{figure}

\subsection{Black widows vs. Redbacks}
\subsubsection{Companion Masses}
\label{sec:compmasses}
In order to compare the companion mass distribution of black widows versus redbacks, we first collated the black widow companion masses. The most accurate way to directly measure the companion mass in a MSP is through the relativistic Shapiro delay \citep{Shapiro64}, which is typically only possible in nearly edge-on systems with precise and long-term pulsar timing solutions \citep[e.g.,][]{Cromartie20}. The significant radio eclipses and orbital variability observed in most black widows typically make them poor targets for long-term timing.
Therefore, typically the best way to infer companion masses in these systems is from the pulsar orbital parameters ($P_{\rm{orb}}$ and $a\,\rm{sin}\,i$) in conjunction with light curve modeling of the companion to constrain the inclination, despite the substantial uncertainties associated with the latter measurements.

For most black widows, a lower limit on the mass of the companion is available via pulsar timing, assuming an edge-on inclination ($i=90^{\circ}$) and a neutron star mass of $1.4\,M_{\odot}$. We list these lower limits in Table~\ref{table:BWcatalog4}. If no inclination constraints are available from optical light curve modeling, we adopt the companion mass assuming $i=60^{\circ}$ (hereafter referred to as the ``median'' mass), since for randomly aligned orbits the systems should be uniformly distributed on the sky in cos($i$). For the systems with modeled optical light curves, we assume the best-fit companion mass derived from these models. 

These median or best-fit (if available) masses are listed in Table~\ref{table:BWcatalog4}. Owing to the large systematic uncertainties associated with deriving inclinations from black widow light curves, we do not include formal uncertainties on these mass estimates.

We assumed the black widow companion star masses were drawn from a normal distribution, and we modeled this sample using a Bayesian MCMC model. We find a median mass $M_{c} = 0.027 \pm 0.003\,M_{\odot}$ with $\sigma=0.015 \pm 0.002\,M_{\odot}$. This can be compared to the distribution of redback companion masses, which have a median mass $M_{c} = 0.39\pm0.05\,M_{\odot}$ with $\sigma=0.12\pm0.05$ \citep{Strader19}. We note that since we do not include formal uncertainties on the individual black widow companion mass estimates in this analysis, the observed dispersion quoted for the black widows is likely an overestimate of the intrinsic dispersion.

We fixed the inclination for a large number of the systems, so as a check on how sensitive these conclusions are to our inclination assumptions, we repeated this analysis after randomly assigning inclination values drawn from a prior that is flat in cos($i$) (i.e., assuming random orientations), finding a median mass $M_{c} = 0.029 \pm 0.004\,M_{\odot}$ with $\sigma=0.021 \pm 0.003\,M_{\odot}$, very similar to our original analysis but with a larger spread. Finally, we analyzed the distribution for only the 17 black widow systems with mass estimates from modeling the photometry, finding $M_{c} = 0.033 \pm 0.004\,M_{\odot}$ with $\sigma=0.016 \pm 0.003\,M_{\odot}$.

We plot the orbital period and median (or best-fit, if available) companion masses for the field black widows and redbacks in Figure~\ref{fig:spiderfig}. We also show the white dwarf companions to field MSPs (blue circles) along with a binary evolution model that shows the well-known relation between the final period and white dwarf mass for systems with a range of initial orbital periods \citep[black line,][]{Tauris99}.

Despite the continued discovery of several new black widow and redback systems in the past few years, the mass distributions appear bimodal and there is a noticeable absence of sources in the companion mass range $\sim0.07-0.1\,M_{\odot}$.

\subsubsection{Orbital Periods}
Similar to our comparison of the masses in the previous section,
we also analyzed the orbital period distributions of the black widows and redbacks. For both subclasses, we included all confirmed and candidate systems, with the exception of the two ``huntsman'' systems that have giant companions in wide ($\gtrsim$5 d) orbits, which were left out of the redback distribution. For the black widows, we find a median orbital period of $P_{\rm{median}}^{\rm{BW}} = 0.229 \pm 0.032$ d with $\sigma = 0.178 \pm 0.024$ d. For the redbacks, we find the orbital periods are longer in the mean, with $P_{\rm{median}}^{\rm{RB}} = 0.389 \pm 0.056$ d and $\sigma = 0.252 \pm 0.043$ d. Despite the significant difference in the mean orbital period, the distributions overlap substantially, so it is challenging to classify a spider on the basis of orbital period alone except at the shortest periods.

\subsubsection{Emission Lines}
As a class, redback companions are significantly brighter than the black widows, so optical spectroscopic follow-up is easier and is much more complete for the redback population. The initial interpretation of emission lines in the optical spectra of some redbacks (e.g., \citealt{Strader15}) was colored by the existence of transitional millisecond pulsars, which sometimes show emission lines from accretion disks (e.g., \citealt{2014MNRAS.444.3004D}) and which are all redbacks. However, the appearance of similar highly-broadened or double-peaked lines in black widows, as well as more detailed study of emission lines in redbacks, suggests that in most cases the emission lines in optical spectra of spiders are more likely associated with the companion (either directly from the chromosphere or a stellar wind), or via emission from the intrabinary shock itself, rather than an accretion disk.

For example, in the candidate redback 1FGL J0523.5--2529, \citet{Halpern22b} infer the optical flaring and spectral emission features are nonthermal, coming from above the photosphere of the secondary and presumably associated with the companion wind outflow and/or the intrabinary shock. A similar conclusion was reached for the origin of the emission lines in the ``huntsmen'' MSP 1FGL J1417.7--4402 \citep{Swihart18}. As another explanation, \citet{Romani15} attributed the emission lines in the black widow PSR J1311--3430 to thermal emission below the stellar photosphere, likely due to pulsar wind-triggered magnetic reconnection that provides a localized heating source.

Setting aside disk-dominated spectra of transitional MSPs or candidate members of this class, there are 15 redbacks with published optical spectroscopy. Of these, 5 (33\%) show prominent emission lines: 1FGL J0523.5--2529 \citep{Halpern22b}, 3FGL J0838.8--2829 \citep{Halpern17b}, PSR J1048+2339 \citep{Strader19}, PSR J1306--40 \citep{Swihart19}, and PSR J1628--3205 \citep{Strader19}. Both of the huntsman sources 1FGL J1417.7--4407 \citep{Strader15,Swihart18} and 2FGL J0846.0+2820 \citep{Swihart17} also show emission, as does the subgiant--MSP binary PSR J1740-5340A in the globular cluster NGC 6397 \citep{Sabbi2003}.

For the black widows, only 10 systems have published spectra, but of these 6 (60\%) show H and/or He in emission (Table~\ref{table:BWcatalog4}). Although the statistics are low, these results suggest that as a class, emission lines are at least as common among black widows as among redbacks, and indeed may be more common. 

If the emission lines are associated with an irradiation-driven wind from the companion or an intrabinary shock, it is not immediately clear whether this should be stronger in redbacks or black widows. Owing to their larger relative Roche Lobe radii, redbacks intercept a larger fraction of emission than do black widows from a source centered on the pulsar. But if the high-energy emission responsible for the optical emission lines is due primarily to X-rays mediated by the intrabinary shock, rather than $\gamma$-rays, the situation becomes more complex. If the shock wraps around the companion in black widows, but around the pulsar in redbacks (e.g., \citealt{Romani16b, Wadiasingh17}), then the weaker shock in black widows might be compensated for by a closer location to the secondary.

To add to the complexity, the appearance of emission lines in these systems is not always stable and predictable. For example, in the redback 3FGL J0838.8--2829 \citep{Halpern17b} and the black widow 4FGL J1408.6--2917 (this work), the emission lines come and go irregularly on short ($\sim$minutes to hours) timescales, with varying strength and phenomenology (i.e., sometimes double-peaked, sometimes single-peaked), and have no clear trend with orbital phase. Other systems, like the huntsman 1FGL J1417.7--4407 \citep{Swihart18}, show persistent, orbital phase-dependent emission that is consistent in data taken over timespans of years. Furthermore, a range of temperatures and/or compositions in the emission line-producing regions is implied by the fact that some systems show only Balmer emission in their spectra (e.g., PSR J1306--40; \citealt{Swihart19}) while others also display prominent He I lines that are occasionally double-peaked (e.g., PSR J1311--3430; \citealt{Romani15}).

Given the large range of phenomenologies, it is possible that the exact origin of the emission lines differ from one system to the next. Future spectroscopic monitoring of new and existing spider MSPs to determine the connections between emission features and the properties of the companion and the intrabinary shock is needed.

\subsubsection{X-ray Emission}
\label{sec:xray_emission}
\begin{figure}[t!]
\begin{center}
	\includegraphics[width=\linewidth]{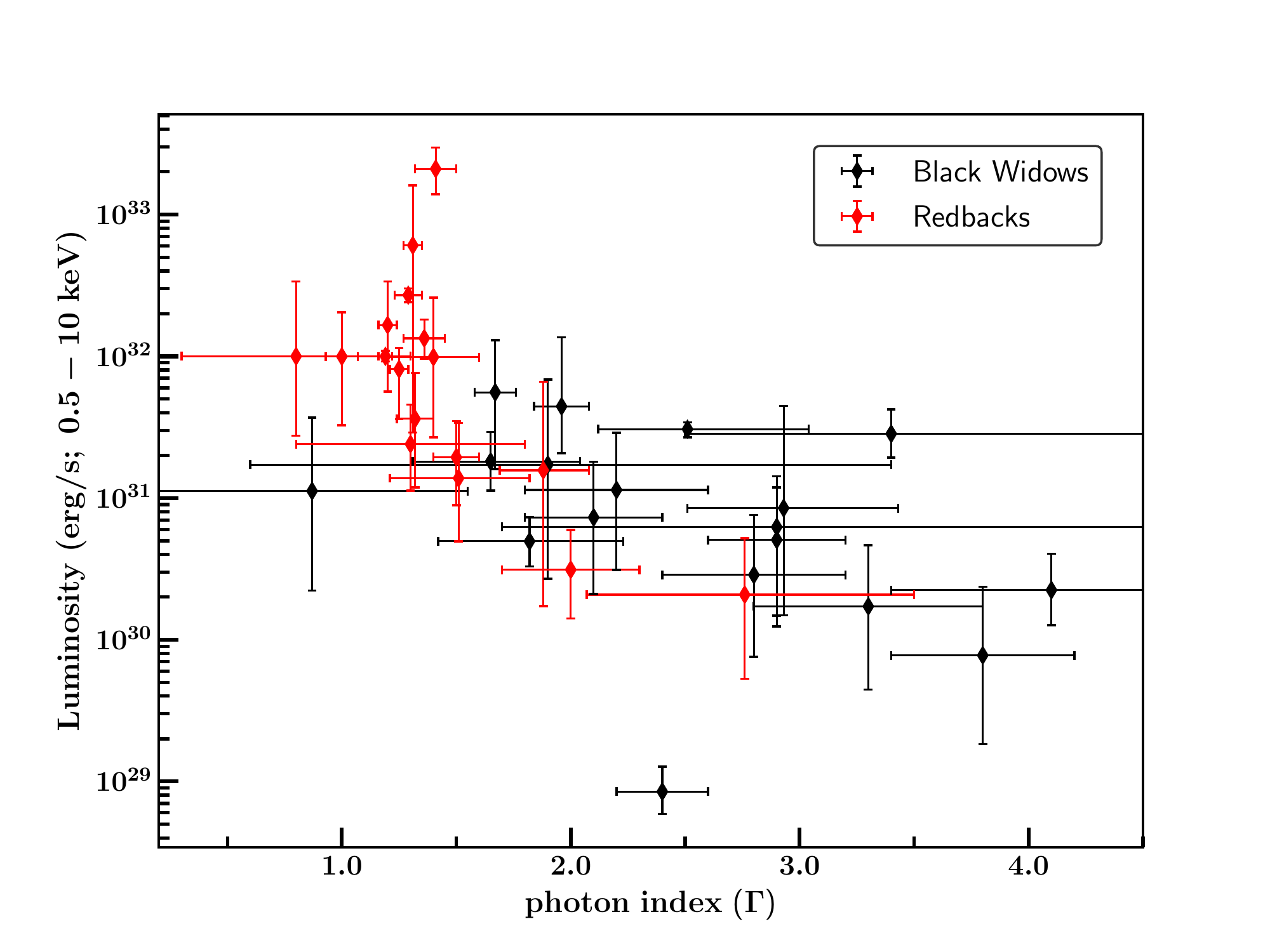}
    \caption{Photon index versus the 0.5--10 keV X-ray luminosity for redbacks (red) and black widows (black). Luminosities were derived as described in Sec.~\ref{sec:measprops}. X-ray emission from redbacks tends to be brighter and harder than in black widows, likely due to more efficient acceleration/compression of the intrabinary shock (see text).}
    \label{fig:LX_gamma}
\end{center}
\end{figure}

We compare the 0.5--10 keV X-ray luminosities ($L_{X}$) and best-fit photon indices ($\Gamma$) for the redbacks and black widows in Figure~\ref{fig:LX_gamma}. The X-ray properties of the redbacks were taken from the cited references in \citet{Lee18} or \citet{Strader19} and scaled to 0.5--10 keV as described in Sec.~\ref{sec:measprops}. Redback X-ray luminosities were derived assuming the distances from \citet{Strader19}.

The average properties of the redback sample are $L_{X} = 2.3\times10^{32}\,\rm{erg\;s^{-1}}$ and $\Gamma = 1.44$, while for the black widows $L_{X} = 1.4\times10^{31}\,\rm{erg\;s^{-1}}$ and $\Gamma = 2.51$. \citet{Lee18} found that X-ray emission in redbacks is brighter and harder than in black widows, although their sample only consisted of X-ray properties from 12 black widows and 8 redbacks. With our larger sample (18 black widows and 17 redbacks/candidates), we again observe evidence for redbacks being brighter and harder than black widows, but also confirm a clear trend in the spider distribution as a whole, with the softer sources typically being intrinsically fainter in X-rays with a nearly continuous distribution spanning over three orders of magnitude in X-ray luminosity.

\begin{figure}[t!]
\begin{center}
	\includegraphics[width=\linewidth]{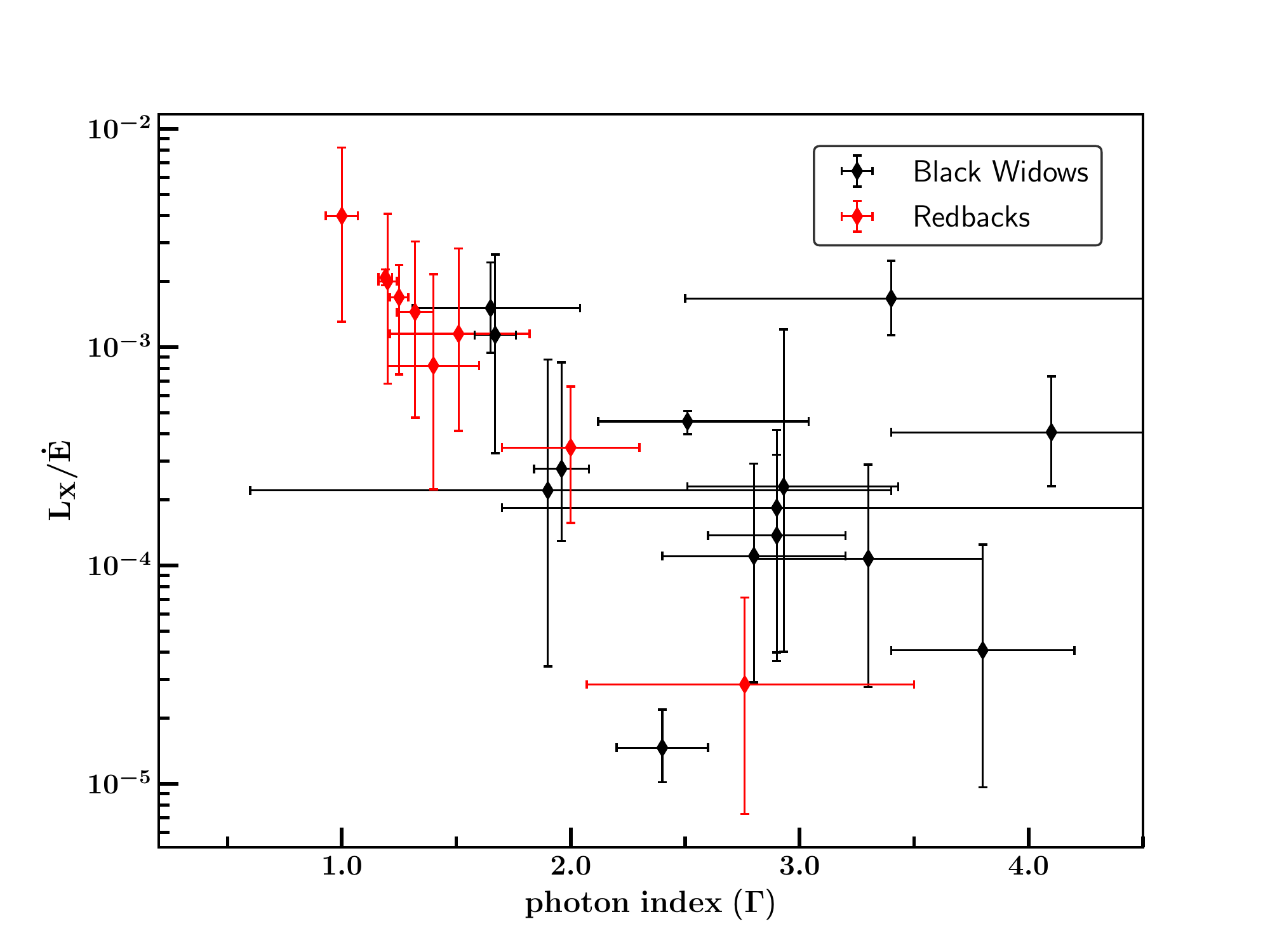}
    \caption{Photon index versus the X-ray luminosity as a fraction of the pulsar spin-down power $\dot{E}$. In general, redbacks convert a larger fraction of their pulsar spin-down power to X-rays. The few black widows with redback-like X-ray properties suggests that in lieu of stellar winds some of these pulsar companions have very strong magnetospheres (see Sec.~\ref{sec:xray_emission}). Despite the low statistics, a rough trend appears that suggests systems with softer X-ray spectra also convert less of their pulsar spin-down power to X-rays.}
    \label{fig:Edot_gamma}
\end{center}
\end{figure}

In Figure~\ref{fig:Edot_gamma} we show the relation between the X-ray photon index and the X-ray luminosity as a fraction of the pulsar spin-down power $\dot{E}$. Although the statistics are limited, 
this figure suggests that, in general, the X-ray luminosity in redbacks represents a much larger percentage of the pulsar spin-down power than in black widows. This figure also shows an apparent trend, whereby systems with a softer X-ray spectrum also convert a smaller fraction of the pulsar spin-down power into X-rays. 

The simplest explanation for these figures is that the intrabinary shock acceleration is vastly more efficient in redbacks. From the perspective of the pulsar, the solid angle subtended by the companion is significantly larger in redbacks than in the smaller black widows. The interplay between the pulsar and companion's wind or magnetosphere then controls the overall geometry of the shock. In general, if the companion's pressure dominates over the pulsar wind, then the shock standoff radius will move farther from the companion and the shock will wrap around the pulsar, and vice-versa if the pulsar wind dominates \citep[e.g.,][]{Romani16b, Wadiasingh17}. In redbacks with well-sampled phase-resolved X-ray light curves, the shock is almost always wrapped around the pulsar, implying stronger companion winds or magnetopheres in these systems compared to the black widows \citep{Wadiasingh18,Merwe20}. As such, in redbacks, a larger fraction of the pulsar's $\dot{E}$ is captured by the shock, naturally resulting in higher intrinsic non-thermal X-ray luminosities.

More interestingly, the results in Figures~\ref{fig:LX_gamma} and \ref{fig:Edot_gamma} for the spectral index suggest that the particle acceleration is also systematically more efficient in redbacks. The pulsars themselves influence the injection spectrum of electron/positron pairs into the shock. Since the pulsars are not intrinsically very different in redbacks from black widows this implies the systematic differences in the spectral index and particle acceleration efficiency results from some conditions at the shock that must differ between the two classes. The spectrum of accelerated particles at the shock is expected to be modified by geometric differences of compression of the pulsar wind's stripes, the local shock magnetic obliquity, and the resultant shock-driven reconnection \citep[e.g.,][]{Summerlin12}.

Close inspection of Figure~\ref{fig:Edot_gamma} shows there are at least two black widows with redback-like $\Gamma$ and $L_{X} / \dot{E}$. These are PSR J1311--3430 and PSR J1653--0158, which share many observed and intrinsic properties. Among the confirmed black widows in our sample, these two systems have the shortest orbital periods. Both systems also have very low-mass H-depleted companions, show prominent emission lines in their optical spectra, and are two of the brightest high Galactic latitude \emph{Fermi} sources. In addition, both their light curves show evidence of nonthermal flux that dominates the optical light near minimum brightness, implying a strong evaporating wind.

One explanation may be that these two black widows have redback-like shocks that wrap around the pulsar instead of around the companion. Since the extremely low-mass companions in black widows likely do not support strong winds, this argues for very strong companion magnetospheres in some spider systems in order to achieve a pressure balance with the relativistic pulsar wind \citep{Wadiasingh18}; some recent observations support the existence of these strong companion magnetospheres \citep[e.g.,][]{Li22}. 

We note there are selection biases involved when analyzing these systems as a population. Typical follow-up observations of \emph{Fermi} sources with \emph{Swift} are a few ksec in duration, reaching X-ray flux limits of a few $\times\,10^{-14}$ erg\;s$^{-1}$\,cm$^{-2}$, so very low-luminosity or more distant systems cannot be detected in X-rays.

Among the 18 confirmed black widows detected in X-rays and that also have a distance estimate listed in Table~\ref{table:BWcatalog2}, the median distance is 2.20 kpc. This can be compared with the median distance of the 19 non-detected sources, which is 2.24 kpc, suggesting that larger distances may not be the primary factor leading to the non-detections.

Even for many of the nearby sources, follow-up observations have just simply not gone deep enough to detect the expected faint X-ray emission. For example, two black widows that are very nearby and that have reliable radio timing parallax distances are PSR J2322--2650 and PSR J2234+0944, which have distances of $0.23^{+0.09}_{-0.05}$ kpc and $0.8^{+0.3}_{-0.2}$ kpc, respectively. PSR J2322--2650 was observed with \emph{Swift} for $\sim$4.7 ksec and has a 0.5--10 keV flux upper limit of $< 1.0\times 10^{-13}$ erg\;s$^{-1}$\,cm$^{-2}$, assuming a power law with $\Gamma=2.5$. Using its parallax distance, this flux corresponds to an  X-ray luminosity upper limit $L_X < 6.4^{+6.0}_{-2.5}\times10^{29}$ erg\;s$^{-1}$. An upper limit for PSR J2234+0944 using similar Swift data corresponds to a luminosity limit $L_X < 4.3^{+3.8}_{-1.9}\times10^{30}$ erg\;s$^{-1}$. These limits are near the bottom envelope of, but not far outside, the range of luminosities currently observed for black widows (Figure~\ref{fig:LX_gamma}).

With future deeper and more sensitive X-ray observations, it will be interesting to see if the tentative correlations seen in Figure~\ref{fig:LX_gamma} and Figure~\ref{fig:Edot_gamma} hold at the lowest X-ray luminosities. Since new systems have recently been discovered with distances $\lesssim$3 kpc, the census of these binaries is clearly far from complete, especially at larger distances, and future multiwavelength programs to find new spider MSPs both inside and outside of $\gamma$-ray source regions could reveal a wider range of phenomenologies.

\subsection{Neutron Star Masses in Black Widows}
\label{sec:NSmasses}
Despite the complex systematic effects involved in modeling the heated companions of black widows, measuring the radial velocity of the secondary with optical spectroscopy provides valuable dynamical constraints that allow for estimates on the mass of the neutron star when coupled with light curve fitting and pulsar timing models.

To date, the most massive neutron star with a precise and direct measurement comes from the relativistic Shapiro delay pulsar timing measurement from the neutron star--white dwarf binary PSR J0740+6620 \citep{Fonseca21,2020NatAs...4...72C}, which has a mass of $M_{\rm{NS}} = 2.08\pm0.07\,M_{\odot}$. There are claims of neutron star masses from some black widows (and at least one redback) that are higher than this value. Perhaps the most notable of these is PSR J0952--0607, the fastest spinning black widow in our sample, which was recently reported to have the most massive well-measured neutron star mass to date at $M_{\rm{NS}} = 2.35 \pm 0.17\,M_{\odot}$ \citep{Romani22b}. Other black widows that have been claimed to host neutron stars $\gtrsim 2.1\,M_{\odot}$ include the original black widow PSR B1957+20 ($2.40\pm0.12\,M_{\odot}$; \citealt{2011ApJ...728...95V}), PSR J1311--3430 ($2.2\pm0.4\,M_{\odot}$; \citealt{Romani15}), PSR J1653--0158 ($2.17\pm0.2\,M_{\odot}$; \citealt{Nieder20}), and PSR J1810+1744 ($2.13\pm0.04\,M_{\odot}$; \citealt{Romani21}). To these can be added the highly-irradiated redback PSR J2215+5135, which has a claimed mass as high as $2.27\pm0.16\,M_{\odot}$ (\citealt{2018ApJ...859...54L,2020ApJ...892..101K}, though see also \citealt{2020MNRAS.499.1758V}).

Despite the increasing sophistication of the light curve modeling in the most recent papers cited above, which include an improved treatment of gravity darkening and non-standard heat transport across the surface of the companion, it has not been established that these models accurately incorporate all relevant physical effects. Given, for example, that an orbital inclination change of only $1.3^{\circ}$ is the difference between a derived mass of $2.1\,M_{\odot}$ and $2.2\,M_{\odot}$ at values around the median inclination of $60^{\circ}$, it is clear that the conclusions drawn about the most massive neutron stars in spiders are affected by even small systemic uncertainties in light curve modeling. In this context, it is relevant that the highest-mass neutron stars have all been found in highly irradiated spiders (black widows or the strongly irradiated redback PSR J2215+5135), rather than in comparably recycled neutron stars in redbacks that have light curves less affected by irradiation, which all have inferred masses $\lesssim 2.1\,M_{\odot}$ \citep{Strader19}.

While it is unclear the extent to which specific individual measurements of black widow neutron star masses are fully reliable, a more secure claim, made already in many of the papers cited above, is that the black widow neutron star mass distribution has a median value significantly larger than the canonical $1.4\,M_{\odot}$; a similar result was found for the redbacks \citep{Strader19}. This is consistent either with these neutron stars having been born massive or having accreted a substantial amount of mass. Given the fast spins and long predicted accretion lifetimes of these binaries as discussed above, accretion seems likely to have played a substantial, if not dominant role.

\section{Discussion and Conclusions}
\label{sec:discussion}

\subsection{The Properties of Black Widows}
In the context of our discovery of a likely new black widow associated with 4FGL J1408.6--2917, we compiled the properties of known black widows. J1408 is now one of~\numBWsTotal~confirmed and candidate black widows in the Galactic field. We showed that the spider companion mass distribution is still strongly bimodal, split between the lower-mass black widows and the more massive redbacks. We also find the orbital periods of the black widows are slightly shorter than for redbacks. Optical emission lines are common in both systems, and although the statistics are limited, they are seen more frequently in the spectra of black widows.

We compared the X-ray properties of the spiders showing that the harder and more luminous X-rays in redbacks implies that acceleration/compression of the intrabinary shock is more efficient in these systems. We also observe the broad relation that spider binaries with harder X-ray spectra tend to convert a larger fraction of the pulsar spin-down power to X-rays. 
Some black widows show spectral indices and $L_{X} / \dot{E}$ values that are comparable to redbacks. If these black widows are unable to power significant winds due to their low masses, an implication could be that the companion magnetospheres in at least some spider binaries are the dominant source of pressure balance supporting the intrabinary shock from the companion side, rather than a wind from the companion.

\subsection{The Origin of Black Widows}
As mentioned above, one of the most puzzling observational findings is that of a bimodal companion mass distribution among spider MSPs, with additional emerging evidence for systematic differences in their X-ray properties. 

This bimodality does not emerge naturally from binary evolution models. Most spiders are expected to have evovled from close binaries where the secondary filled its Roche Lobe on the main sequence or early in its post-main sequence evolution, recycling the neutron star. Compared to the well-studied cataclysmic variables with white dwarf primaries, the subsequent evolution is strongly affected by irradiation by the pulsar wind, which continues even when accretion has ceased. This can occur ``naturally" if magnetic braking shuts off when the donor becomes fully convective or perhaps even earlier in the evolution, if accretion-induced irradation leads to a Roche Lobe underfilling donor (e.g., \citealt{ Benvenuto12,Chen13,DeVito20}). In either case, the proximate cause for the companion to become a redback is a high level of irradiation, leading to faster evaporation, compared to the black widow case.

However, in the more recent models from \citet{Ginzburg20a}, evaporation alone cannot cause significant enough mass loss to explain the observed spider populations. Instead, the irradation is proposed to change the internal structure of the companion, allowing it to maintain a strong magnetic field even down to very low masses. The irradition-driven evaporative wind couples to this magnetic field and the companion can maintain stable Roche-lobe overflow for a much longer time and at longer orbital periods, giving more efficient magnetic braking and evolutionary timescales that agree better with the observed spider population \citep{Ginzburg21}.

In this model, the two parameters that control the evolution are the MeV $\gamma$-ray pulsar luminosity ($L_{\rm{MeV}}$) that evaporates the companion, and the broader spectrum irradiating $\gamma$-ray luminosity ($L_{\rm{irr}} > L_{\rm{MeV}}$) that is deposited in the companion atmosphere. $L_{\rm{irr}}$ lengthens the thermal timescale of the companion, allowing it to maintain Roche-lobe overflow at longer orbital periods as described above. In this model, different values of $L_{\rm{irr}}$ give a range of observed periods, while the mass gap between the two populations is proposed to originate from a bimodal distribution of companion magnetic fields (weaker for redbacks, stronger for black widows). 

A key prediction of this model is that $L_{\rm{irr}}$ is correlated with the orbital period of the binary ($L_{\rm{irr}} \propto P_{\rm{orb}}^{2.5}$). \citet{Ginzburg21} found that the measured pulsar spin-down luminosity does not agree with this relation, but that the high-energy $\gamma$-ray luminosities from \emph{Fermi} do (see their Figure 12), providing support for this theoretical model.

\begin{figure}[]
\begin{center}
	\includegraphics[width=\linewidth]{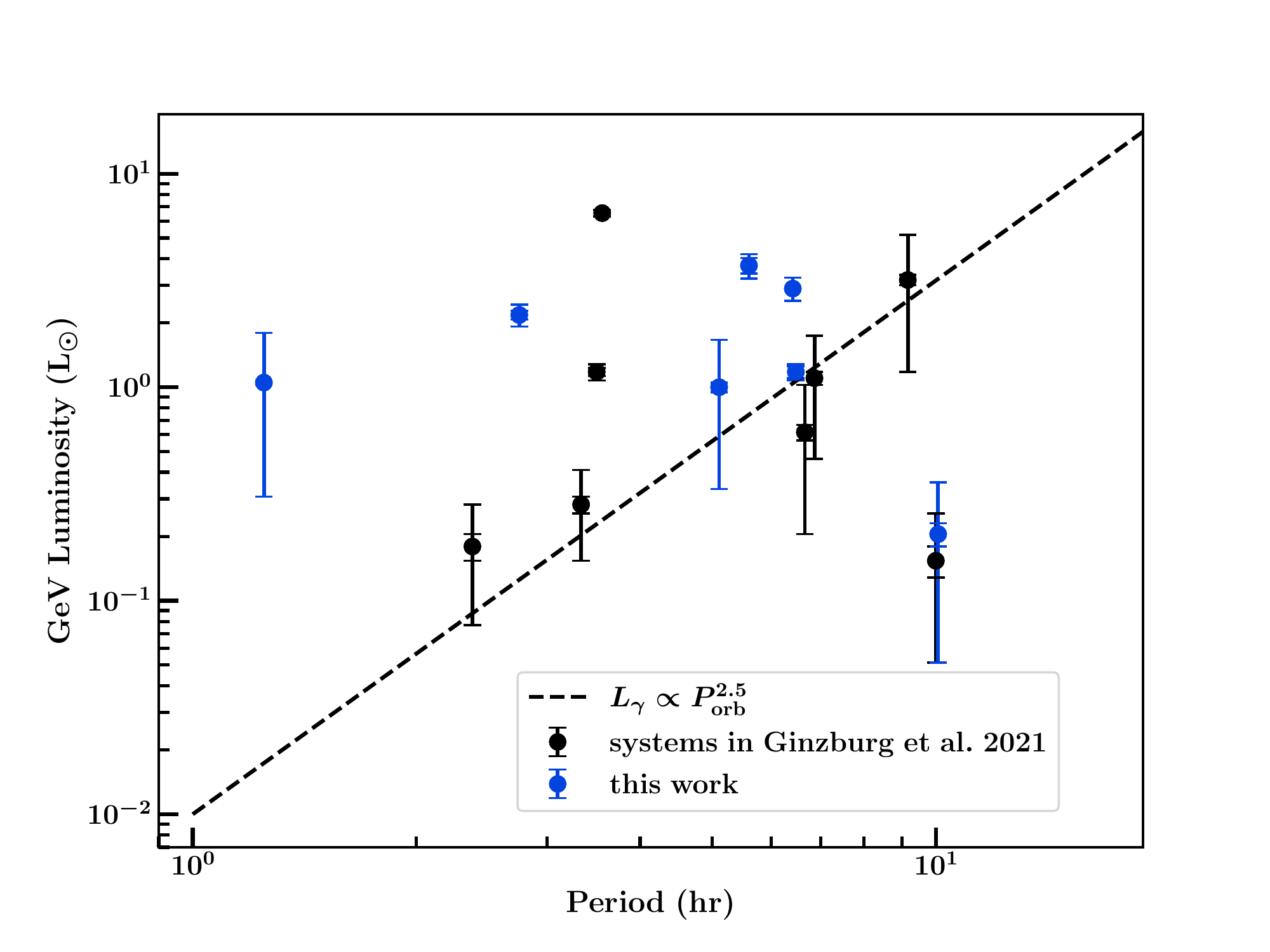}
    \caption{The $\gamma$-ray luminosity of black widows does not clearly correlate with orbital period. The figure shows 
    the 0.1--100 GeV \emph{Fermi} $\gamma$-ray luminosity versus the binary orbital period, adapted from Figure 12 in \citet{Ginzburg21}. Black points are the systems plotted in the original figure with updated values for $L_{\gamma}$ from Table~\ref{table:BWcatalog3}, supplemented by 7 additional systems (blue points) with reliable distance estimates (see text). The two sets of error bars on each point represent the uncertainty associated with the flux and distance, respectively. Evolutionary models from \citet{Ginzburg21} predict the pulsar’s irradiating luminosity, potentially tracked by the \emph{Fermi} GeV luminosity, is correlated with the orbital period (dashed line). With our updated luminosities and additional systems, we find the data do not fit the predicted relation.}
    \label{fig:Ginzfig}
\end{center}

\end{figure}

We have revisited this proposed relation, adding an additional 7 systems that have well-constrained distance estimates from either a parallax measurement or from light curve models and updating the data for the others where relevant (Figure~\ref{fig:Ginzfig}). With the updated data, the observational support for this predicted relation is weaker. For most systems the $\gamma$-ray luminosity provides sufficient energy to power the theoretically inferred $L_{\rm{irr}}$, with all but two systems lying above the dashed line in Figure~\ref{fig:Ginzfig}. But, unlike in \citet{Ginzburg21}, we see no clear relation between $L_{\gamma}$ and orbital period. In the context of this model, this finding is consistent with the idea $L_{\gamma}$ is a poor proxy for the irradiating luminosity $L_{\rm{irr}}$. Alternatively, the fact that almost all the observations lie above the predicted line might result from inefficient heat transport between the two hemispheres of the tidally locked black widow companion. According to \citet{Ginzburg21}, it is the fraction of $L_{\rm{irr}}$ that is transported to the companion's non-irradiated nightside which sets its thermal timescale and therefore correlates with the orbital period. The lack of relation between $L_{\gamma}$ and $P_{\rm{orb}}$ could also potentially be explained by a beaming or another efficiency factor, leading to a model that is more akin to the assumptions made in some previous models (e.g., \citealt{Benvenuto12,Chen13}).

Beaming, or more generally a variation in the irradiation efficiency, may also be relevant for understanding the relationship between black widows and the related class of accreting millisecond X-ray pulsars (AMXPs). Some AMXPs, such as SAX J1808.4--3658, have very similar MSP spin periods, orbital periods, and companion masses to black widows (e.g., \citealt{Patruno21}). One possibility is that AMXPs come from initially similar systems to black widows but have a lesser or minimal degree of MSP irradiation, leading to lower mass loss rates and shorter orbital periods as observed (e.g., \citealt{Ginzburg21,DAntona22}). The difference in orbital periods indicates that AMXPs and black widows cannot be exactly the same population of systems at different points in a cyclical accretion cycle. However,  in principle a difference in irradiation efficiency among individual systems could have been present either early in their evolution, or could have developed over time. At least for some systems, an evolutionary connection is also possible: many binary evolution models of the SAX J1808.4--3658 suggest that this system will eventually stop accreting and become a detached black widow \citep[e.g.,][]{Chakrabarty98, Ergma99, Chen17, Tailo18, Goodwin20}

In any case, the present work emphasizes the continuing evidence for a bimodal spider companion mass distribution: not only the existence of redbacks, but also the ``gap" in companion masses in the range $0.07$--$0.1\,M_{\odot}$. Notably, longer orbital period AMXPs with hydrogen-rich donor stars also appear to avoid this mass gap \citep{Patruno21,DAntona22}, consistent with a common physical cause that must be addressed by a successful model.

\subsection{Concluding Remarks}

The explosion in the number of spider MSPs found by \emph{Fermi} and associated follow-up was seen as notable even from earliest discoveries (e.g., \citealt{Ray12,Roberts13}). Subsequent work has shown that these spiders are not simply a curious ``add-on" to the existing subpopulations of MSPs: spiders are now relatively common, making up $\gtrsim 15-20$\% of the fully recycled MSPs with known companion classifications. In addition, they represent a dominant fraction (70--80\%) of the fastest spinning field MSPs, some of which may also be among the most massive neutron stars known. Additional observational and theoretical work to understand the formation and evolution of black widows and redbacks is an extremely promising route to understand the extremes of neutron star behavior.

\section*{Acknowledgements}
Special thanks to Z. Wadiasingh for thoughtful discussion on the X-ray emission and to S. Ginzburg for helpful insights on the evolutionary models.

This research was performed while SJS held a NRC Research Associateship award at the Naval Research Laboratory. Work at the Naval Research Laboratory is supported by NASA DPR S-15633-Y.

We also acknowledge support
from NSF grants AST-1714825 and AST-2205550 and the Packard Foundation.

Based on observations obtained at the Southern Astrophysical Research (SOAR) telescope, which is a joint project of the Minist\'{e}rio da Ci\^{e}ncia, Tecnologia, Inova\c{c}\~{o}es e Comunica\c{c}\~{o}es (MCTIC) do Brasil, the U.S. National Optical Astronomy Observatory (NOAO), the University of North Carolina at Chapel Hill (UNC), and Michigan State University (MSU).

We acknowledge the use of public data from the Swift data archive.

The Green Bank Observatory is a facility of the National Science Foundation operated under cooperative agreement by Associated Universities, Inc

This work used observations obtained with XMM–Newton, an ESA science mission with instruments and contributions directly funded by ESA Member States and NASA.

This research has made use of data and software provided by the High Energy Astrophysics Science Archive Research Center (HEASARC), which is a service of the Astrophysics Science Division at NASA/GSFC and the High Energy Astrophysics Division of the Smithsonian Astrophysical Observatory.

This work has made use of data from the European Space Agency (ESA) mission {\it Gaia} (\url{https://www.cosmos.esa.int/gaia}), processed by the {\it Gaia} Data Processing and Analysis Consortium (DPAC, \url{https://www.cosmos.esa.int/web/gaia/dpac/consortium}). Funding for the DPAC has been provided by national institutions, in particular the institutions participating in the {\it Gaia} Multilateral Agreement.

Based on observations obtained at the international Gemini Observatory, a program of NSF’s NOIRLab, which is managed by the Association of Universities for Research in Astronomy (AURA) under a cooperative agreement with the National Science Foundation on behalf of the Gemini Observatory partnership: the National Science Foundation (United States), National Research Council (Canada), Agencia Nacional de Investigaci\'{o}n y Desarrollo (Chile), Ministerio de Ciencia, Tecnolog\'{i}a e Innovaci\'{o}n (Argentina), Minist\'{e}rio da Ci\^{e}ncia, Tecnologia, Inova\c{c}\~{o}es e Comunica\c{c}\~{o}es (Brazil), and Korea Astronomy and Space Science Institute (Republic of Korea).

\begin{deluxetable*}{llllllll}
\setlength\tabcolsep{3.5pt}
\tabletypesize{\scriptsize}
\tablecaption{Black widow Catalog}
\tablehead{
\colhead{ID} & \colhead{Other ID} & \colhead{R.A. (ICRS)\tablenotemark{\scriptsize{a}}} & \colhead{Decl. (ICRS)} & \colhead{$P_{\rm{spin}}$} & \colhead{$\dot{P_{\rm{obs}}}$} & \colhead{$a\,\rm{sin}\,i$} & \colhead{$\dot{E}$}\\[-7pt]
\colhead{} & \colhead{} & \colhead{(h:m:s)} & \colhead{($^\circ:\arcmin:\arcsec$)} & \colhead{(ms)} & \colhead{($10^{-20}$ s)} & \colhead{(lt-s)} & \colhead{($10^{34}$ erg s$^{-1}$)}
}
\startdata
PSR J0023+0923 & 4FGL J0023.4+0920 & 00:23:16.877498(8) & +09:23:23.8604(3) & 3.050203104754390(4) & 1.142345(14) & 0.03484136(5) & 1.6\\[-3pt]
PSR J0251+2606 & 4FGL J0251.0+2605 & 02:51:02.5537(5) & +26:06:09.97(2) & 2.5415543469461(3) & 0.7572(8) & 0.065681(8) & 1.8\\[-3pt]
PSR J0312--0921 & 4FGL J0312.1--0921 & 03:12:06.2 & --09:21:56 & 3.704 & 1.972 & \nodata & 1.5\\[-3pt]
PSR J0610--2100 & 4FGL J0610.2--2100 & 06:10:13.595462(17) & --21:00:27.9313(4) & 3.8613247042986(6) & 1.235(6) & 0.0734891(4) & 0.85\\[-3pt]
PSR J0636+5129 & PSR J0636+5128 & 06:36:04.847464(5) & +51:28:59.96547(11) & 2.868952846616078(8) & 0.34483(7) & 0.00898636(6) & 0.58\\[-3pt]
PSR J0952--0607 & 4FGL J0952.1--0607 & 09:52:08.32141(5) & --06:07:23.490(2) & 1.41379835502312(12) & 0.4773(8) & 0.0626670(9) & 6.7\\[-3pt]
PSR J1124--3653 & 4FGL J1124.0--3653 & 11:24:01.116(3) & --36:53:19.10(4) & 2.41 & \nodata & \nodata & \nodata\\[-3pt]
PSR J1301+0833 & 4FGL J1301.6+0834 & 13:01:38.26 & +08:33:57.5 & 1.84 & \nodata & \nodata & \nodata\\[-3pt]
PSR J1311--3430 & 4FGL J1311.7--3430 & 13:11:45.7242(2) & --34:30:30.350(4) & 2.5603710316720(3) & 2.0964(14) & 0.010581(4) & 4.9\\[-3pt]
PSR J1446--4701 & 4FGL J1446.6--4701 & 14:46:35.712054(8) & --47:01:26.78210(14) & 2.194695780881595(15) & 0.98075(3) & 0.06401226(15) & 3.7\\[-3pt]
PSR J1513--2550 & 4FGL J1513.4--2549 & 15:13:23.32059(6) & --25:50:31.285(3) & 2.1190675651177(1) & 2.161(2) & 0.0408132(7) & 9.0\\[-3pt]
PSR J1544+4937 & 4FGL J1544.0+4939 & 15:44:04.48722(2) & +49:37:55.2545(2) & 2.15928839043292(5) & 0.2933(5) & 0.0328680(4) & 1.2\\[-3pt]
PSR J1555--2908 & 4FGL J1555.7--2908 & 15:55:40.6586(10 & --29:08:28.426(13) & 1.78750176696926(20) & 4.45502(4) & 0.1514468(1) & 31.0\\[-3pt]
PSR J1641+8049 & 4FGL J1641.2+8049 & 16:41:20.8381(4) & +80:49:52.9142(8) & 2.02117938468221(9) & 0.895(9) & 0.0640793(3) & 4.3\\[-3pt]
PSR J1653--0158 & 4FGL J1653.6--0158 & 16:53:38.05381(5) & --01:58:36.8930(5) & 1.9676820247057(2) & 0.2402(3) & 0.01071(1) & 1.2\\[-3pt]
PSR J1719--1438 &  & 17:19:10.07293(5) & --14:38:00.942(4) & 5.7901517700238(5) & 0.8043(7) & 0.0018212(7) & 0.16\\[-3pt]
PSR J1720--0533 & & 17:19:55 & --05:30:05 & 3.27 & \nodata & \nodata & \nodata\\[-3pt]
PSR J1731--1847 & 4FGL J1731.7--1850 & 17:31:17.609823(17) & --18:47:32.666(3) & 2.34455954688568(11) & 2.5407(5) & 0.1201611(6) & 7.8\\[-3pt]
PSR J1745+1017 & 4FGL J1745.5+1017 & 17:45:33.8371(7) & +10:17:52.523(2) & 2.6521296710897(4) & 0.2729(15) & 0.088172(1) & 0.58\\[-3pt]
PSR J1745--2324 & PSR J1745--23 & 17:45:30(24) & -23:25(7) & 5.41669986(14) & \nodata & 0.06247(6) & \nodata\\[-3pt]
PSR J1805+0615 & 4FGL J1805.6+0615 & 18:05:42.39969(3) & +06:15:18.606(13) & 2.1289064590218(5) & 2.2758(9) & 0.087728(15) & 9.3\\[-3pt]
PSR J1810+1744 & 4FGL J1810.5+1744 & 18:10:37.28(1) & +17:44:37.38(7) & 1.66 & \nodata & 0.095 & \nodata\\[-3pt]
PSR J1833--3840 & 4FGL J1833.0--3840 & 18:33:04.6 & --38:40:46 & 1.87 & 1.773 & \nodata & 11.0\\[-3pt]
PSR J1928+1245 &  & 19:28:45.39360(6) & +12:45:53.374(3) & 3.0216063479651(6) & 1.680(10) & 0.018951(1) & 2.4\\[-3pt]
PSR J1946--5403 & 4FGL J1946.5--5402 & 19:46:34.497(3) & --54:03:42.51(4) & 2.710 & \nodata & 0.0435 & \nodata\\[-3pt]
PSR J1959+2048 & 4FGL J1959.5+2048 & 19:59:36.76988(5) & +20:48:15.1222(6) & 1.60740168480632(3) & 1.68515(9) & 0.0892253(6) & 16.0\\[-3pt]
 & PSR B1957+20\\[-3pt]
PSR J2017--1614 & 4FGL J2017.7--1612 & 20:17:46.1478(8) & --16:14:15.51(5) & 2.3142872649224(4) & 0.245(5) & 0.043655(5) & 0.78\\[-3pt]
PSR J2047+1053 & 4FGL J2047.3+1051 & 20:47:10.246(3) & +10:53:07.80(4) & 4.29 & \nodata & \nodata & \nodata\\[-3pt]
PSR J2051--0827 & 4FGL J2051.0--0826 & 20:51:07.519768(18) & --08:27:37.7497(8) & 4.50864182000643(11) & 1.2733(7) & 0.0450720(3) & 0.55\\[-3pt]
PSR J2052+1219 & 4FGL J2052.7+1218 & 20:52:47.77803(15) & +12:19:59.022(5) & 1.98525628181868(8) & 0.67037(20) & 0.061377(4) & 3.4\\[-3pt]
PSR J2055+3829 &  & 20:55:10.306550(4) & +38:29:30.90571(6) & 2.08929030219107(3) & 0.09996(5) & 0.0452618(2) & 0.43\\[-3pt]
PSR J2115+5448 & 4FGL J2115.1+5449 & 21:15:11.7678(1) & +54:48:45.154(2) & 2.602876738872(2) & 7.49(1) & 0.044846(1) & 17.0\\[-3pt]
PSR J2214+3000 & 4FGL J2214.6+3000 & 22:14:38.853711(10) & +30:00:38.19160(14) & 3.119226581323024(12) & 1.47285(4) & 0.0590813(3) & 1.9\\[-3pt]
PSR J2234+0944 & 4FGL J2234.7+0943 & 22:34:46.854176(7) & +09:44:30.2224(3) & 3.627027895734199(12) & 2.00998(6) & 0.06842966(13) & 1.7\\[-3pt]
PSR J2241--5236 & 4FGL J2241.7--5236 & 22:41:42.0269841(10) & --52:36:36.239590(11) & 2.1866997725548446(10) & 0.689656(3) & 0.025795324(11) & 2.6\\[-3pt]
PSR J2256--1024 & 4FGL J2256.8--1024 & 22:56:56.39294(7) & --10:24:34.385(3) & 2.29453181696499(3) & 1.13535(10) & 0.08296575(5) & 3.7\\[-3pt]
PSR J2322--2650 &  & 23:22:34.64004(3) & --26:50:58.3171(6) & 3.46309917908790(11) & 0.05834(15) & 0.0027849(6) & 0.055\\[-3pt]
4FGL J0336.0+7502 &  & 03:36:10.1811 & +75:03:17.268 & \nodata & \nodata & \nodata & \nodata\\[-3pt]
4FGL J0935.3+0901 & & 09:35:20.719 & +09:00:35.90 & \nodata & \nodata & \nodata & \nodata\\[-3pt]
ZTF J1406+1222 & & 14:06:56.173(4) & +12:22:43.398(3) & \nodata & \nodata & \nodata & \nodata\\[-3pt]
4FGL J1408.6--2917 &  & 14:08:26.789 & --29:22:21.212 & \nodata & \nodata & \nodata & \nodata
\enddata
\label{table:BWcatalog}
\tablenotetext{a}{Coordinates are taken from the ATNF pulsar database if pulsations have been detected, otherwise the best position of the optical counterpart is used.}
\end{deluxetable*}


\begin{deluxetable*}{lcccccl}[!t]
\tablecaption{Black widow Catalog}
\tablehead{
\colhead{ID} & \colhead{DM} & \colhead{CL02\tablenotemark{\scriptsize{a}}} & \colhead{Y17\tablenotemark{\scriptsize{b}}} & \colhead{Other dist.\tablenotemark{\scriptsize{c}}} & \colhead{References} & \colhead{Final Dist.\tablenotemark{\scriptsize{d}}}\\[-8pt]
\colhead{} & \colhead{pc cm$^{-3}$} & \colhead{(kpc)} & \colhead{(kpc)} & \colhead{(kpc)}& \colhead{} & \colhead{(kpc)}
}
\startdata
J0023+0923 & 14.3 & 0.69 & 1.25 & 1.1(2); 2.23(8) & (1); (2) & 1.1(2)\\
J0251+2606 & 20.2 & 0.82 & 1.17 & 3.26(10) & (2) & 1.2(4)\\
J0312--0921 & 20.5 & 0.87 & 0.82 & \nodata & \nodata & 0.8(2)\\
J0610--2100 & 60.7 & 3.54 & 3.26 & $2.24^{+0.70}_{-0.57}$ & (3) & $2.24^{+0.70}_{-0.57}$\\
J0636+5129 & 11.1 & 0.49 & 0.21 & $1.1^{+0.6}_{-0.3}$; 1.05(1) & (1); (2) & $1.1^{+0.6}_{-0.3}$\\
J0952--0607 & 22.4 & 0.97 & 1.74 & $6.26^{+0.36}_{-0.40}$ & (57) & $6.26^{+0.36}_{-0.40}$\\
J1124--3653 & 44.9 & 1.72 & 0.99 & $2.72^{+0.10}_{-0.08}$ & (2) & 1.0(3)\\
J1301+0833 & 13.2 & 0.67 & 1.23 & $2.23^{+0.08}_{-0.13}$ & (2) & $2.23^{+0.08}_{-0.13}$\\
J1311--3430 & 37.8 & 1.41 & 2.43 & 2.6 & (4) & 2.4(7)\\
J1446--4701 & 55.8 & 1.46 & 1.57 & \nodata & \nodata & 1.6(5)\\
J1513--2550 & 46.9 & 1.95 & 3.96 & \nodata & \nodata & 4.0 $\pm$ 1.2\\
J1544+4937 & 23.2 & 1.23 & 2.99 & 2.0--5.0 & (5) & 3.0(9)\\
J1555--2908 & 75.9 & 2.65 & 7.56 & 5.1(2) & (6) & 5.1(2)\\
J1641+8049 & 31.1 & 1.65 & 3.04 & \nodata & \nodata & 3.0(9)\\
J1653--0158 & \nodata & \nodata & \nodata & 0.84(40); $1.00^{+1.31}_{-0.46}$ & (7); (8) & $1.00^{+1.31}_{-0.46}$\\
J1719--1438 & 36.9 & 1.21 & 0.34 & \nodata & \nodata & 0.3(1)\\
J1720--0533 & 36.8 & 1.35 & 0.19 & \nodata & \nodata & 0.2(1)\\
J1731--1847 & 106.5 & 2.55 & 4.78 & \nodata & \nodata & 4.8 $\pm$ 1.4\\
J1745+1017 & 244.9 & 1.26 & 1.21 & \nodata & \nodata & 1.2(4)\\
J1745--2324 & 24.0 & 4.48 & 7.94 & \nodata & \nodata & 7.9$\pm$2.4\\
J1805+0615 & 64.9 & 2.48 & 3.89 & \nodata & \nodata & 3.9$\pm$1.2\\
J1810+1744 & 39.7 & 2.00 & 2.36 & 3.03(1) & (9) & 3.03(1)\\
J1833--3840 & 78.6 & 2.05 & 4.65 & \nodata & \nodata & 4.7$\pm$1.4\\
J1928+1245 & 179.2 & 6.08 & 6.08 & \nodata & \nodata & 6.1$\pm$1.8\\
J1946--5403 & 23.7 & 0.87 & 1.15 & \nodata & \nodata & 1.2(4)\\
J1959+2048 & 29.1 & 2.49 & 1.73 & $2.22^{+0.03}_{-0.02}$;$2.57^{+1.84}_{-0.77}$ & (10); (11) & $2.57^{+1.84}_{-0.77}$\\
J2017--1614 & 25.4 & 1.10 & 1.44 & \nodata & \nodata & 1.4(4)\\
J2047+1053 & 34.6 & 2.05 & 2.79 & \nodata & \nodata & 2.8(8)\\
J2051--0827 & 20.7 & 1.04 & 1.47 & 2.5(2) & (58) & 2.5(2)\\
J2052+1219 & 42.0 & 2.44 & 3.92 & 3.94(7) & (2) & 3.94(7)\\
J2055+3829 & 91.8 & 4.36 & 4.59 & \nodata & \nodata & 4.6$\pm$1.4\\
J2115+5448 & 77.4 & 3.39 & 3.11 & \nodata & \nodata & 3.1(9)\\
J2214+3000 & 22.5 & 1.54 & 1.67 & $0.4^{+0.2}_{-0.1}$ & (1) & $0.4^{+0.2}_{-0.1}$\\
J2234+0944 & 17.8 & 1.00 & 1.59 & $0.8^{+0.3}_{-0.2}$ & (1) & $0.8^{+0.3}_{-0.2}$\\
J2241--5236 & 11.4 & 0.51 & 0.96 & $1.24^{+0.04}_{-0.05}$ & (2) & $1.24^{+0.04}_{-0.05}$\\
J2256--1024 & 13.8 & 0.65 & 1.33 & 2.0(6) & (12) & 2.0(6)\\
J2322--2650 & 6.1 & 0.32 & 0.76 & $0.23^{+0.09}_{-0.05}$ & (13) & $0.23^{+0.09}_{-0.05}$\\
J0336.0+7502 & \nodata & \nodata & \nodata & \nodata & \nodata\\
J0935.3+0901 & \nodata & \nodata & \nodata & \nodata & \nodata\\
J1406+1222 & \nodata & \nodata & \nodata & 1.14(20) & (28) & 1.14(20)\\
J1408.6--2917 & \nodata & \nodata & \nodata & \nodata & \nodata
\enddata
\label{table:BWcatalog2}
\tablenotetext{a}{Distance using the pulsar dispersion measure and the \citet{Cordes02} electron density model.}
\tablenotetext{b}{Distance using the pulsar dispersion measure and the \citet{Yao17} electron density model.}
\tablenotetext{c}{Parallax or light curve derived distance estimate.}
\tablenotetext{d}{Final adopted distance.}
\end{deluxetable*}


\begin{deluxetable*}{lcccccc}[!t]
\tablecaption{Black widow Catalog}
\tablehead{
\colhead{ID} & \colhead{$F_{\gamma}$\tablenotemark{\scriptsize{a}}} & \colhead{$L_{\gamma}$} & \colhead{$F_{X}$\tablenotemark{\scriptsize{b}}} & \colhead{$\Gamma$} & \colhead{References\tablenotemark{\scriptsize{c}}} & \colhead{$L_{X}$}\\[-8pt]
\colhead{} & \colhead{($10^{-12}$erg\;s$^{-1}$\,cm$^{-2}$)} & \colhead{($10^{33}$ erg s$^{-1}$)} & \colhead{($10^{-14}$erg\;s$^{-1}$\,cm$^{-2}$)} & \colhead{} & \colhead{} & \colhead{($10^{30}$ erg s$^{-1}$)}
}
\startdata
J0023+0923 & 7.8(6) & 1.1(1) & $4.6^{+1.6}_{-1.1}$ & 3.3~$\pm$~0.5 & (18) & $1.72^{+2.92}_{-1.28}$\\
J0251+2606 & 4.9(4) & 0.8(1) & \nodata & \nodata & \nodata & \nodata\\
J0312--0921 & 5.5(4) & 0.4(1) & \nodata & \nodata & \nodata & \nodata\\
J0610--2100 & 7.2(5) & 4.3(3) & \nodata & \nodata & \nodata & \nodata\\
J0636+5129 & \nodata & \nodata & $1.61^{+0.24}_{-0.24}$ & 2.4~$\pm$~0.2 & (19) & $0.08^{+0.04}_{-0.03}$\\
J0952--0607 & 2.4(3) & 11.3~$\pm$~1.4 & 0.652 & $2.51^{+0.53}_{-0.39}$ & (20) & $30.6^{+3.6}_{-3.8}$\\
J1124--3653 & 12.5(6) & 1.5(1) & $5.5^{+1.3}_{-1.0}$ & 2.1~$\pm$~0.3 & (21) & $7.3^{+10.7}_{-5.2}$\\
J1301+0833 & 7.7(5) & 4.6(3) & \nodata & \nodata & \nodata & \nodata\\
J1311--3430 & 60.6~$\pm$~1.2 & 41.8(8) & $6.04^{+0.28}_{-0.27}$ & $1.67^{+0.09}_{-0.09}$ & (22) & $55.7^{+74.3}_{-39.7}$\\
J1446--4701 & 7.7(7) & 2.4(2) & $1.9^{+2.2}_{-0.8}$ & $2.93^{+0.50}_{-0.42}$ & (18) & $8.5^{+36.1}_{-7.0}$\\
J1513--2550 & 7.6(6) & 14.6~$\pm$~1.1 & \nodata & \nodata & \nodata & \nodata\\
J1544+4937 & 2.4(3) & 2.5(3) & \nodata & \nodata & \nodata & \nodata\\
J1555--2908 & 4.7(6) & 14.5~$\pm$~1.9 & \nodata & \nodata & \nodata & \nodata\\
J1641+8049 & 2.0(3) & 2.1(3) & \nodata & \nodata & \nodata & \nodata\\
J1653--0158 & 34.3~$\pm$~1.0 & 4.1(1) & $21.5^{+10.2}_{-6.8}$ & $1.65^{+0.39}_{-0.34}$ & (23); (24) & $18.1^{+11.2}_{-6.8}$\\
J1719--1438 & \nodata & \nodata & \nodata & \nodata & \nodata & \nodata\\
J1720--0533 & \nodata & \nodata & \nodata & \nodata & \nodata & \nodata\\
J1731--1847 & 5.2~$\pm$~1.1 & 14.2~$\pm$~3.0 & $0.63^{+0.62}_{-0.34}$ & $1.9^{+1.5}_{-1.3}$ & (18) & $17.2^{+51.3}_{-14.5}$\\
J1745+1017 & 7.6(6) & 1.3(1) & \nodata & \nodata & \nodata & \nodata\\
J1745--2324 & \nodata & \nodata & \nodata & \nodata & \nodata & \nodata\\
J1805+0615 & 5.3(5) & 9.6~$\pm$~1.0 & \nodata & \nodata & \nodata & \nodata\\
J1810+1744 & 23.2(9) & 25.5(9) & $1.72^{+0.44}_{-0.35}$ & 2.2~$\pm$~0.4 & (18) & $11.4^{+17.4}_{-8.3}$\\
J1833--3840 & 2.8(5) & 7.5~$\pm$~1.2 & \nodata & \nodata & \nodata & \nodata\\
J1928+1245 & \nodata & \nodata & \nodata & \nodata & \nodata & \nodata\\
J1946--5403 & 9.8(5) & 1.7(1) & $2.87^{+1.39}_{-0.96}$ & $1.82^{+0.41}_{-0.40}$ & (27) & $4.95^{+2.39}_{-1.66}$\\
J1959+2048 & 15.7(9) & 12.4(7) & $5.60^{+0.24}_{-0.26}$ & 1.96~$\pm$~0.12 & (25) & $44.3^{+91.7}_{-23.6}$\\
J2017--1614 & 6.5(6) & 1.5(1) & \nodata & \nodata & \nodata & \nodata\\
J2047+1053 & 4.3(6) & 4.0(5) & $1.2^{+0.8}_{-0.5}$ & 0.87~$\pm$~0.68 & (18) & $11.2^{+25.7}_{-9.0}$\\
J2051--0827 & 2.5(3) & 1.9(4) & $0.30^{+0.24}_{-0.13}$ & 4.1~$\pm$~0.7 & (18) & $2.24^{+1.80}_{-0.97}$\\
J2052+1219 & 4.6(6) & 8.5~$\pm$~1.0 & $0.34^{+0.43}_{-0.27}$ & $2.9^{+1.6}_{-1.2}$ & (27) & $6.24^{+7.96}_{-5.00}$\\
J2055+3829 & \nodata & \nodata & \nodata & \nodata & \nodata & \nodata\\
J2115+5448 & 7.0(7) & 8.1(8) & $2.5^{+1.2}_{-0.8}$ & $3.4^{+1.1}_{-0.9}$ & (27) & $28.4^{+13.8}_{-9.1}$\\
J2214+3000 & 32.6(7) & 0.6(1) & $1.81^{+0.93}_{-0.56}$ & 3.8~$\pm$~0.4 & (18) & $0.78^{+1.59}_{-0.59}$\\
J2234+0944 & 10.0(6) & 0.8(1) & \nodata & \nodata & \nodata & \nodata\\
J2241--5236 & 25.0~$\pm$~1.1 & 4.6(2) & $2.60^{+0.83}_{-0.58}$ & 2.8~$\pm$~0.4 & (26) & $2.87^{+4.72}_{-2.11}$\\
J2256--1024 & 8.2(5) & 3.9(2) & $2.39^{+0.42}_{-0.36}$ & 2.9~$\pm$~0.3 & (18) & $5.07^{+6.83}_{-3.59}$\\
J2322--2650 & \nodata & \nodata & \nodata & \nodata & \nodata & \nodata\\
J0336.0+7502 & 8.08(52) & \nodata & \nodata & \nodata & \nodata & \nodata\\
J0935.3+0901 & 4.56(54) & \nodata & 12.75~$\pm$~1.32 & $1.88^{+0.25}_{-0.22}$ & (31) & \nodata\\
J1406+1222 & \nodata & \nodata & \nodata & \nodata & \nodata & \nodata\\
J1408.6--2917 & 5.21(69) & \nodata & $2.4^{+1.6}_{-1.1}$ & $1.66^{+0.55}_{-0.48}$ & (27) & \nodata
\enddata
\tablenotetext{\scriptsize{a}}{$\gamma$-ray flux from \emph{Fermi}-LAT over the energy range 0.1--100 GeV \citep{4FGLDR3}.}
\tablenotetext{\scriptsize{b}}{Unabsorbed 0.5--10 keV X-ray flux.}
\tablenotetext{\scriptsize{c}}{Reference for X-ray properties.}
\label{table:BWcatalog3}
\end{deluxetable*}

\begin{deluxetable*}{llccccccc}[!t]
\tablecaption{Black widow Catalog}
\tablehead{
\colhead{ID} & \colhead{Orbital Period} & \colhead{Modeled} & \colhead{Spectroscopy?} & \colhead{$M_{c}^{\rm{min}}$\tablenotemark{\scriptsize{a}}} & \colhead{$M_{c}$\tablenotemark{\scriptsize{b}}} & \colhead{References} & \colhead{Discovery} & \colhead{References}\\[-8pt]
\colhead{} & \colhead{(days)} & \colhead{Photometry?} & \colhead{Emission Lines (Y/N)} & \colhead{($M_{\odot}$)} & \colhead{($M_{\odot}$)} & \colhead{} & \colhead{} & \colhead{}
}
\startdata
J0023+0923 & 0.13879914382(4) & \checkmark & -- & 0.016 & 0.018 & (2) & Radio follow-up of \emph{Fermi} & (32)\\
J0251+2606 & 0.2024406403(9) & \checkmark & -- & 0.024 & 0.032 & (2) & Radio follow-up of \emph{Fermi} & (33)\\
J0312--0921 & 0.0975 & -- & -- & 0.009 & 0.010 & \nodata & Radio follow-up of \emph{Fermi} & (55)\\
J0610--2100 & 0.2860160068(6) & \checkmark & -- & 0.021 & 0.022 & (3) & Parkes survey & (34)\\
J0636+5129 & 0.066551340763(16) & \checkmark & -- & 0.007 & 0.018 & (2) & GBT survey & (35)\\
J0952--0607 & 0.2674610347(5) & \checkmark & \checkmark(N) & 0.019 & 0.026; 0.032 & (14); (57) & Radio follow-up of \emph{Fermi} & (36)\\
J1124--3653 & 0.2291666 & \checkmark & -- & \nodata & 0.041 & (2) & Radio follow-up of \emph{Fermi} & (32)\\
J1301+0833 & 0.27 & \checkmark & \checkmark(N) & \nodata & 0.045 & (2) & Radio follow-up of \emph{Fermi} & (37)\\
J1311--3430 & 0.0651157335(7) & \checkmark & \checkmark(Y) & 0.008 & 0.0104 & (4) & Opt/X-ray search of \emph{Fermi} & (38)\\
J1446--4701 & 0.27766607699(15) & -- & -- & 0.019 & 0.022 & \nodata & Parkes survey & (39)\\
J1513--2550 & 0.1786354505(8) & -- & -- & 0.016 & 0.019 & \nodata & Radio follow-up of \emph{Fermi} & (40)\\
J1544+4937 & 0.1207729895(1) & \checkmark & -- & 0.017 & 0.025 & (5) & Radio follow-up of \emph{Fermi} & (41)\\
J1555--2908 & 0.23350026854(11) & \checkmark & \checkmark(N) & 0.051 & $0.060^{+0.005}_{-0.003}$ & (6) & Radio follow-up of \emph{Fermi} & (42)\\
J1641+8049 & 0.0908739634(1) & -- & -- & 0.040 & 0.047 & \nodata & GBT survey & (35)\\
J1653--0158 & 0.0519447575(4) & \checkmark & \checkmark(Y) & 0.010 & 0.013 & (7) & Opt/X-ray search of \emph{Fermi} & (24); (43)\\
J1719--1438 & 0.0907062900(12) & -- & -- & 0.0011 & 0.0013 & \nodata & Parkes survey & (44)\\
J1720--0533 & 0.131666 & -- & -- & \nodata & \nodata & \nodata & FAST survey & (56)\\
J1731--1847 & 0.3111341185(10) & -- & -- & 0.033 & 0.039 & \nodata & Parkes survey & (45)\\
J1745+1017 & 0.730241444(1) & -- & -- & 0.014 & 0.016 & \nodata & Radio follow-up of \emph{Fermi} & (46)\\
J1745--2324 & 0.165562(10) & -- & -- & 0.027 & 0.030 & \nodata & Parkes survey & (47)\\
J1805+0615 & 0.3368720310(48) & -- & -- & 0.023 & 0.027 & \nodata & Radio follow-up of \emph{Fermi} & (33)\\
J1810+1744 & 0.14817083 & \checkmark & \checkmark(N) & 0.043 & 0.065 & (17); (9) & Radio follow-up of \emph{Fermi} & (32)\\
J1833--3840 & 0.900 & -- & -- & \nodata & \nodata & \nodata & Parkes survey & \nodata\\
J1928+1245 & 0.1366347269(8) & -- & -- & 0.009 & 0.010 & \nodata & Arecibo survey & (48)\\
J1946--5403 & 0.130 & -- & -- & 0.021 & 0.025 & \nodata & Radio follow-up of \emph{Fermi} & (49)\\
J1959+2048 & 0.3819666069(8) & \checkmark & \checkmark(Y) & 0.021 & 0.036 & (2) & Arecibo survey & (50)\\
J2017--1614 & 0.0978252578(4) & -- & -- & 0.026 & 0.030 & \nodata & Radio follow-up of \emph{Fermi} & (40)\\
J2047+1053 & 0.12 & -- & -- & 0.036 & 0.042 & \nodata & Radio follow-up of \emph{Fermi} & (37)\\
J2051--0827 & 0.09911025490(4) & \checkmark & -- & 0.027 & $0.039^{+0.010}_{-0.011}$ & (58) & Parkes survey & (51)\\
J2052+1219 & 0.1146136251(2) & \checkmark & -- & 0.033 & 0.042 & (2) & Radio follow-up of \emph{Fermi} & (33)\\
J2055+3829 & 0.12959037294(1) & -- & -- & 0.022 & 0.027 & (16) & NRT survey & (16) \\
J2115+5448 & 0.135322188(3) & -- & -- & 0.022 & 0.025 & \nodata & Radio follow-up of \emph{Fermi} & (40)\\
J2214+3000 & 0.41663294591(20) & -- & -- & 0.013 & 0.015 & \nodata & Radio follow-up of \emph{Fermi} & (52)\\
J2234+0944 & 0.41966003706(17) & -- & -- & 0.015 & 0.018 & \nodata & Radio follow-up of \emph{Fermi} & (37)\\
J2241--5236 & 0.14567224025(2) & \checkmark & -- & 0.012 & 0.016 & (2) & Radio follow-up of \emph{Fermi} & (53)\\
J2256--1024 & 0.21288263050(7) & \checkmark & -- & 0.030 & 0.032 & (17) & Radio follow-up of \emph{Fermi} & (32)\\
J2322--2650 & 0.322963997(6) & -- & -- & 0.00074 & 0.00086 & \nodata & Parkes survey & (13)\\
J0336.0+7502 & 0.15492408(38) & \checkmark & -- & \nodata & \nodata & (29) & Opt/X-ray search of \emph{Fermi} & (29)\\
J0935.3+0901 & 0.10153276(36) & \checkmark & \checkmark(Y) & \nodata & \nodata & (30) & Opt/X-ray search of \emph{Fermi} & (54)\\
J1406+1222 & 0.043056621(2) & \checkmark & \checkmark(Y) & \nodata & \nodata & (28) & Optical survey & (28)\\
J1408.6--2917 & 0.14261385(150) & \checkmark & \checkmark(Y) & \nodata & \nodata & (27) & Opt/X-ray search of \emph{Fermi} & (27)\\
\enddata
\tablenotetext{\scriptsize{a}}{Minimum companion mass assuming $i=90^{\circ}$ and a neutron star mass of $1.4\,M_{\odot}$.}
\tablenotetext{\scriptsize{b}}{Best-fit companion mass from modeling the optical photometry. If no light curve exists we adopt the median companion mass assuming $i=60^{\circ}$.}
\tablenotetext{}{{\bf References.} (1) \citet{Nano11}; (2) \citet{Draghis19}; (3) \citet{Wateren22}; (4) \citet{Romani15}; (5) \citet{Tang14}; (6) \citet{Kennedy22}; (7) \citet{Nieder20}; (8) \citet{GaiaDR3}; (9) \citet{Romani21}; (10) \citet{Kandel20}; (11) \citet{Romani22}; (12) \citet{Crowter20}; (13) \citet{Spiewak18}; (14) \citet{Nieder19}; (15) \citet{Stappers99}; (16) \citet{Guillemot19}; (17) \citet{Breton13}; (18) \citet{Arumugasamy15}; (19) \citet{Spiewak16}; (20) \citet{Ho19}; (21) \citet{Gentile14}; (22) \citet{An17}; (23) \citet{Cheung12}; (24) \citet{Romani14}; (25) \citet{Huang12}; (26) \citet{An18}; (27) this work; (28) \citet{Burdge22}; (29) \citet{Li21}; (30) \citet{Halpern22}; (31) \citet{Zheng22}; (32) \citet{Hessels11}; (33) \citet{Cromartie16}; (34) \citet{Burgay06}; (35) \citet{Stovall14}; (36) \citet{Bassa17b}; (37) \citet{Ray12}; (38) \citet{Romani12}; (39) \citet{Keith12}; (40) \citet{Sanpa16}; (41) \citet{Bhattacharyya13}; (42) \citet{Ray22}; (43) \citet{Kong14}; (44) \citet{Bailes11}; (45) \citet{Bates11}; (46) \citet{Barr13}; (47) \citet{Cameron20}; (48) \citet{Parent19}; (49) \citet{Camilo15}; (50) \citet{Fruchter88}; (51) \citet{Stappers96}; (52) \citet{Ransom11b}; (53) \citet{Keith11}; (54) \citet{Wang20}; (55) \citet{Tabassum21}; (56) \citet{WangS21}; (57) \citet{Romani22b}; (58) \citet{Dhillon22}
}
\label{table:BWcatalog4}
\end{deluxetable*}

\clearpage
\bibliography{main}

\appendix

\section{An Unrelated AGN}
\label{sec:AGN}

Here we discuss the properties of the other X-ray source 
found within the error ellipse of 4FGL J1408.6--2917. Based on the results presented below, we classify this source as an AGN that is unrelated to the $\gamma$-ray source.

In the \emph{Swift} observations described in Sec.~\ref{sec:Swiftobs}, this source was detected with $\sim$8 counts, corresponding to a 0.3--10 keV flux of $1.0^{+2.0}_{-0.8}\times10^{-13}$ erg cm$^{\rm{-2}}$ s$^{\rm{-1}}$. In our \emph{XMM} observation of this region, this source is also present, with an unabsorbed 0.3--10 keV flux of $1.5^{+0.3}_{-0.3}\times10^{-13}$ erg cm$^{\rm{-2}}$ s$^{\rm{-1}}$. This is marginally brighter than in the \emph{Swift} data, but within the large uncertainties of that shallower data. Fitting the \emph{XMM}/EPIC  spectrum as described in Sec.~\ref{sec:XMMdesrip}, we find the source has a very hard spectrum with $\Gamma = 1.18 \pm 0.18$.

The X-ray source matches spatially to a single candidate optical counterpart, a faint Pan-STARRS \citep{Flewelling20} source with $r'=21.6$, listed as Pan-STARRS ID 72822121474471499. The optical color of this source is nominally blue ($g'-r' = -0.09\pm0.26$) though with a large uncertainty. The source is detected in \emph{WISE} $W1$ and $W2$ filters and has an extremely red Pan-STARRS/WISE color of $r'-W1 = 5.0\pm0.3$ and a WISE color of $W1-W2 = 0.85$. These photometric properties and the hard X-ray photon index are consistent with the classification of this source as a background AGN (e.g., \citealt{Stern12}).

Since this source (Pan-STARRS 72822121474471499) is somewhat closer to the center of the 4FGL error ellipse than the black widow J1408, we considered the possibility that this AGN was instead the counterpart of the \emph{Fermi}-LAT $\gamma$-ray source. This is very unlikely solely based on the existence of the black widow
candidate J1408: such binaries are rare and there is no other variable class that can mimic the short period,  large amplitude, and emission lines we observe. To serendipitously find one within a \emph{Fermi}-LAT error ellipse, but have it not be associated with the GeV source, would be extraordinarily unlikely.

Focusing on the properties of the AGN, it has a lower X-ray flux and softer $\gamma$-ray spectrum than most \emph{Fermi}-detected AGN, and the lack of $\gamma$-ray variability and slight evidence for $\gamma$-ray spectrum curvature also provide additional evidence that the 4FGL source is not associated with an AGN \citep{Ajello20}. Since the high-latitude source density of X-ray sources down to the flux of Pan-STARRS 72822121474471499 is $\sim 10$--20 per deg$^{2}$ (e.g., \citealt{Carrera07}), there is a $\sim 15-30$\% probability of a chance coincidence between a \emph{Fermi} error ellipse the size of that for 4FGL J1408.6--2917 and an unrelated X-ray source of this flux level. Since the bulk of these X-ray sources are background AGN, it is reasonable to expect that many high-latitude \emph{Fermi}-LAT sources will have error ellipses that encompass unrelated background AGN, as appears to be the case for 4FGL J1408.6--2917.

\end{document}